\documentclass[journal]{IEEEtran}
\ifCLASSINFOpdf \else \fi

\usepackage[cmex10]{amsmath}
\usepackage{amssymb}

\usepackage{algorithmic}
\usepackage{array}
\usepackage{url}
\usepackage[noadjust]{cite}

\newtheorem{definition}{Definition}

\newtheorem{proposition}{Proposition}
\newtheorem{remark}{Remark}

\usepackage{fancyhdr}
\usepackage{amsfonts}
\usepackage{xfrac}
\usepackage{color}
\usepackage{pgfplots}           
\pgfplotsset{compat=1.16}
\usepackage{enumerate}

\usepackage{cite}
\usepackage{mathrsfs}
\usepackage{subcaption}
\usepackage{stfloats}

\usepackage{afterpage}
\usepackage{tikz}
\usepackage{mathdots}
\usepackage{yhmath}
\usepackage{cancel}
\usepackage{color}
\usepackage{siunitx}
\usepackage{array}
\usepackage{multirow}
\usepackage{textcomp}
\usepackage{gensymb}
\usepackage{tabularx}
\usepackage{booktabs}
\usepackage{graphicx}
\usepackage{xcolor}

\begin{document}

\title{\huge Zero-Energy Reconfigurable Intelligent Surfaces (zeRIS)}

\author{Dimitrios Tyrovolas,~\IEEEmembership{Graduate Student Member,~IEEE,}
Sotiris A. Tegos,~\IEEEmembership{Member,~IEEE,}\\
Vasilis K. Papanikolaou,~\IEEEmembership{Member,~IEEE,}
Yue Xiao, Prodromos-Vasileios Mekikis,~\IEEEmembership{Member,~IEEE,}\\ Panagiotis D. Diamantoulakis,~\IEEEmembership{Senior Member,~IEEE,} Sotiris Ioannidis,
Christos K. Liaskos,~\IEEEmembership{Member,~IEEE,} \\ and George K. Karagiannidis,~\IEEEmembership{Fellow,~IEEE}
\thanks{D. Tyrovolas is with the 
Department of Electrical and Computer Engineering, Aristotle University of Thessaloniki, 54124 Thessaloniki, Greece, and with the Department of Electrical and Computer Engineering, Technical University of Chania, Chania, Greece (tyrovolas@auth.gr).}  
\thanks{S. A. Tegos and P. D. Diamantoulakis are with the Department of Electrical and Computer Engineering, Aristotle University of Thessaloniki, 54124 Thessaloniki, Greece (tegosoti@auth.gr, padiaman@auth.gr).} 
\thanks{V. K. Papanikolaou is with the Institute for Digital Communications (IDC), Friedrich-Alexander-University Erlangen-Nuremberg (FAU), 91054 Erlangen, Germany (vasilis.papanikolaou@fau.de).} 
\thanks{Y. Xiao is with the School of Information Science and Technology, Southwest Jiaotong University, Chengdu, China (xiaoyue@swjtu.edu.cn).} 
\thanks{P.-V. Mekikis is with Hilti Corporation, Feldkircher Strasse 100, 9494 Schaan, Liechtenstein (akis.mekikis@hilti.com)}
\thanks{S. Ioannidis is with the Department of Electrical and Computer Engineering, Technical University of Chania, Chania, Greece (sotiris@ece.tuc.gr).}
\thanks{C. K. Liaskos is with the Computer Science Engineering Department, University of Ioannina, Ioannina, and Foundation for Research and Technology Hellas (FORTH), Greece (cliaskos@ics.forth.gr).} 
\thanks{G. K. Karagiannidis is with the Department of Electrical and Computer Engineering, Aristotle University of Thessaloniki, 54124 Thessaloniki, Greece and with the Artificial Intelligence and Cyber Systems Research Center, Lebanese American University (LAU), Lebanon (geokarag@auth.gr).}
\thanks{The work has been funded by the European Union’s Horizon 2020 research and innovation programs under grant agreement No 101021659 (SENTINEL) and grant agreement No 958478 (EnerMan). The work of Y. Xiao was supported by the National NSFC 62201477 and
Sichuan Youth Fund Project 2023NSFSC1374} }

\maketitle

\begin{abstract}
A primary objective of the forthcoming sixth generation (6G) of wireless networking is to support demanding applications, while ensuring energy efficiency. Programmable wireless environments (PWEs) have emerged as a promising solution, leveraging reconfigurable intelligent surfaces (RISs), to control wireless propagation and deliver exceptional quality-of-service. In this paper, we analyze the performance of a network supported by \textit{zero-energy RISs (zeRISs)}, which harvest energy for their operation and contribute to the realization of PWEs. Specifically, we investigate joint energy-data rate outage probability and the energy efficiency of a zeRIS-assisted communication system by employing three harvest-and-reflect (HaR) methods, i) power splitting, ii) time switching, and iii) element splitting. Furthermore, we consider two zeRIS deployment strategies, namely BS-side zeRIS and UE-side zeRIS. Simulation results validate the provided analysis and examine which HaR method performs better depending on the zeRIS placement. Finally, valuable insights and conclusions for the performance of zeRIS-assisted wireless networks are drawn from the presented results.
\end{abstract}

\begin{IEEEkeywords}
Reconfigurable Intelligent Surface (RIS), harvest-and-reflect (HaR), Self-sustainable, Zero-Energy Devices (ZEDs), Performance Analysis 
\end{IEEEkeywords}

\IEEEpeerreviewmaketitle

\section{Introduction}

In the pursuit of addressing the stringent requirements of future wireless networks \cite{yue}, the groundbreaking paradigm of programmable wireless environments (PWEs) has emerged to revolutionize the wireless propagation process, by turning it into a software-defined procedure and enabling pervasive connectivity \cite{liaskosmag,liaskos2019network,SRE}. In this direction, a prominent approach to implement PWEs involves integrating reconfigurable intelligent surfaces (RISs) into the environment, which can manipulate power, direction, polarization, and phase of incident waves through their reflective elements in an almost passive manner \cite{Access, venkatesh, alexandropoulos}. \color{black}By employing RISs, networks can orchestrate customized propagation routes, significantly enhance wireless channel quality, and facilitate cutting-edge applications such as intelligent sensing, accurate localization, efficient data transmission, over-the-air computing, and immersive extended reality experiences \cite{tyrovolas2022performance, isac, localization, inforis, IACE, XR}\color{black}. Thus, through the dynamic control of the wireless environment, the future of wireless communication could be reshaped, leading to more adaptive and efficient networks capable of meeting the diverse demands of various applications and user scenarios \cite{liaskosmag,Access,SRE,activeris}. 

To harness the full capabilities of a PWE, which allow it to manipulate transmitted waves, it is imperative to deploy a large number of RISs within the wireless propagation environment. Nevertheless, considering the importance of energy efficiency in the context of 6G networks, the development and adoption of sustainable techniques to achieve this objective is essential \cite{zed}. Wireless power transfer (WPT) facilitates the establishment of an environmental-friendly network based on zero-energy devices (ZEDs), which are powered by harvesting energy from radio-frequency (RF) signals \cite{phd,rui1,zed,vpapanikk}. Therefore, by converting RISs into ZEDs, referred in this paper as \textit{zero-energy RISs (zeRISs)}, we can facilitate the implementation of energy-efficient PWEs and fulfill the principal criterion for sustainability \cite{rui1,vpapanikk,liaskos2019network}. More specifically, by incorporating a suitable energy harvesting (EH) system and capitalizing on the absorption functionality of RISs \cite{liaskosmag}, we can harvest the necessary energy for their functioning and transform conventional RISs into zeRISs, thus eliminating the need for designated power sources. Consequently, zeRISs can be proposed as an important component of future 6G eco-friendly networks, while simultaneously enabling the realization of various PWE-enabled services through their electromagnetic functionalities, such as beam-steering and beam-splitting, among others \cite{liaskosmag,beamform,beamsplit}.

\subsection{State-of-the-art and Motivation}
\color{black}
Recent studies have explored the transformative potential of converting RISs into ZEDs to enhance their performance in an energy-efficient manner \cite{ntontin1,ARES,sris,Yu2023,jamalipour,cunhua,robust,ambient}. For example, the authors of \cite{ntontin1} and \cite{ARES} considered the feasibility of making a conventional RIS self-sustainable, while \cite{sris} examined the WPT efficiency of a network assisted by such a RIS. However, the authors in \cite{sris} concentrated primarily on specific EH methods without addressing the broader impact of factors such as varying propagation conditions, zeRIS locations, or the potential benefits of other EH techniques. Similarly, \cite{Yu2023} analyzed a network in which ZEDs perform simultaneous wireless information and power transfer (SWIPT) through a wireless-powered RIS, with a focus on minimizing transmit power, yet did not investigate alternative EH methods that could offer improved performance. Further research in \cite{jamalipour} introduced a self-sustainable RIS to enhance the performance of a wireless-powered communication network, where a hybrid access point transfers energy to the RIS and multiple users, enabling subsequent information transmission after the EH process. In \cite{cunhua}, the authors tackled the sum-rate maximization problem in a self-sustainable RIS-aided system, wherein information users and WPT clients are served separately. Meanwhile, \cite{robust} aimed to minimize the transmit power of a self-sustainable RIS-assisted system, taking into account the signal-to-noise ratio (SNR) of the receivers and energy budget constraints. Finally, \cite{ambient} proposed an innovative RIS architecture that replaces traditional reflecting elements with backscattering devices, which harvest energy and produce amplitude and phase shift effects on received waves instead of remodulating them. Despite these advances though, there exists no work that has established suitable metrics for quantifying the performance of a zeRIS-assisted network with varying EH methods, propagation conditions, and zeRIS topologies. Addressing this gap is crucial for a comprehensive understanding the key factors that influence the performance of a
zeRIS-assisted network, offering guidance for network designers.\color{black}

\subsection{Contribution}
In this paper, we investigate the performance of a point-to-point communication system assisted by a zeRIS, which harvests the necessary energy for its operation and facilitates information transmission through its beam-steering functionality, utilizing the \emph{harvest-and-reflect (HaR)} protocol. More specifically, the primary contributions of this work can be summarized as follows:

\begin{itemize}
\item \color{black}We explore the performance of three prominent zeRIS-HaR methods: i) power splitting (PS), involving tunable absorption and beam manipulation per zeRIS element; ii) time switching (TS), which alternates between absorption and beam manipulation over time per element; and iii) element splitting (ES), entailing distinct absorption or beam manipulation per element.\color{black}
\item To assess the distinct features of each HaR method, we express the required energy for the zeRIS operation when each method is applied, and derive closed-form expressions for the joint energy-data rate outage probability of the system. This probability quantifies the trade-off between EH and information transmission, which can be employed for the derivation of the energy efficiency for all investigated HaR methods.
\item To characterize the impact of the zeRIS placement on the network's performance, we derive expressions for the joint energy-data rate outage probability and for the system's energy efficiency in two scenarios: the zeRIS is located near the base station (BS), i.e., \textit{BS-side} zeRIS, or near the user equipment (UE), i.e., \textit{UE-side} zeRIS. 
\item We present simulation results to verify the mathematical analysis and evaluate the influence of the available number of reflecting elements, the applied HaR method, the zeRIS location, and the propagation conditions on the performance of a zeRIS-assisted wireless network. Unlike conventional RIS-assisted scenarios, we argue that the performance of \textit{UE-side} zeRIS-assisted systems is asymmetric to that of the \textit{BS-side} case, even when the system parameters remain the same for both deployment scenarios.
\end{itemize}

\subsection{Structure}
The remaining of the paper is organized as follows. The zeRIS architecture, the analyzed system model, as well as the examined HaR methods and zeRIS deployment strategies are described in Section II. Moreover, the performance analysis is given in Section III, while our results are presented in Section IV. Finally, Section V concludes the paper.
\color{black}
Throughout this paper, several specialized mathematical notations and functions are employed to facilitate the analysis. Specifically, $\Gamma(\cdot)$ represents the gamma function, $\gamma(\cdot, \cdot)$ denotes the lower incomplete gamma function, and $I_n(\cdot)$ refers to the n-th order modified Bessel function of the first kind. Finally, $\delta(\cdot)$ represents the Dirac function, and {\small$\mathbb{E}[\cdot]$} denotes the expected value of a random variable.
\color{black}

\section{Zero-Energy Reconfigurable Intelligent Surface (zeRIS)}

\subsection{zeRIS Architecture}

Our main goal is to develop a reconfigurable metasurface unit, i.e., a zeRIS, that is entirely self-sufficient in terms of energy, relying only on incoming waves for power. To achieve this, we suggest a new programmable metasurface architecture, i.e., zeRIS architecture, that includes reflecting elements with adjustable impedance and a specialized controller responsible for the zeRIS connection with the network and the reflecting elements' configuration. Specifically, we assume that each zeRIS element contains an RF-to-DC converter, converting a predefined portion of the incoming wave power into a DC voltage. This voltage is subsequently stored in a capacitor that functions as short-term energy storage, ensuring a steady flow of electricity to the zeRIS controller and the impedance-switching components and, thus, the initial required energy for the zeRIS operation. It should be mentioned that the element design can direct a variable part of the incoming wave to the rectifier without altering the wave's phase that reaches the impedance switch. This can be achieved by adding a non-reciprocal energy-moving component immediately after the metallic element patch, which allows energy to enter and exit a central structure in both symmetric and asymmetric ways \cite{sounas2013giant,newMS}. However, if non-reciprocal components are not preferred due to manufacturing challenges, it can be considered that the rectifier does not affect the accuracy of the impedance switch by designing the components properly. 

With these components, each zeRIS element can alter the amplitude and phase of received waves, following the phased antenna array operating principle and the programming instructions from the zeRIS controller. Regarding the physical structure, each zeRIS element has an antenna pattern layer, which can be made using basic printing techniques, followed by an RF insulator and a ground layer whose thickness and material depend on the desired resonant frequency. Lastly, through-vias can connect the antenna pattern to the element rectifier and the element impedance switch.

\subsection{System Model}
We consider a downlink communication network that consists of: i) a BS equipped with a single-antenna, ii) a zeRIS with $N$ reflecting elements, and iii) a single-antenna UE. Due to the harsh propagation conditions, the BS takes advantage of the zeRIS's ability to steer its impinging radiation towards the UE, assuming that perfect channel state information is available at the BS. Hence, taking into account that the zeRIS needs to harvest energy to operate, it is vital to determine the most appropriate way to perform the absorption functionality without compromising the network's reliability.

\subsection{zeRIS-HaR Methods}
In this work, \color{black}inspired by the Simultaneously Transmitting and Reflecting (STAR) RIS paradigm \cite{starris} the following three primary zeRIS-HaR techniques are examined:\color{black}

\begin{figure}[t]
\centering
\begin{subfigure}{.45\textwidth}
	\centering
 \includegraphics[width=0.97\textwidth]{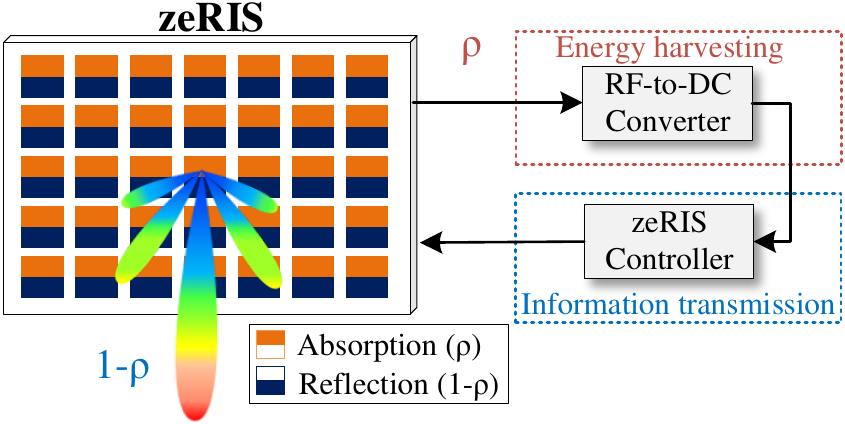}
    \caption{Power splitting (PS)}
    \vspace{3mm}
\end{subfigure}
\begin{subfigure}{.45\textwidth}
	\centering
 \includegraphics[width=0.97\textwidth]{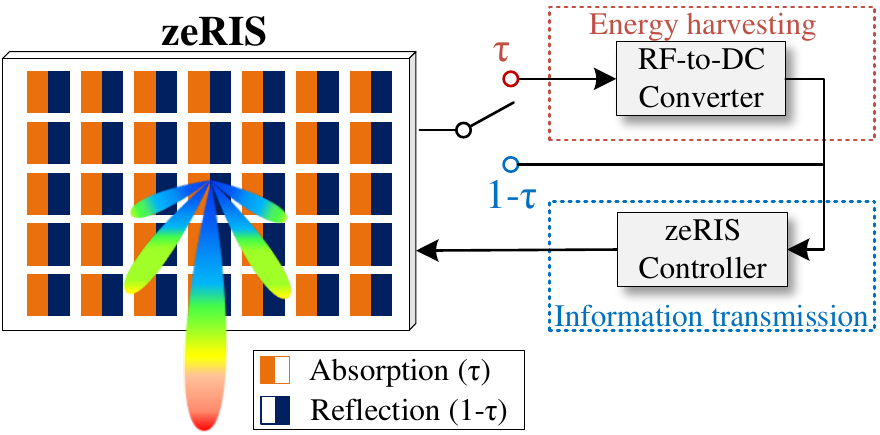}
    \caption{Time switching (TS)}
    \vspace{3mm}
    \end{subfigure}
\begin{subfigure}{.45\textwidth}
	\centering
    \includegraphics[width=0.97\textwidth]{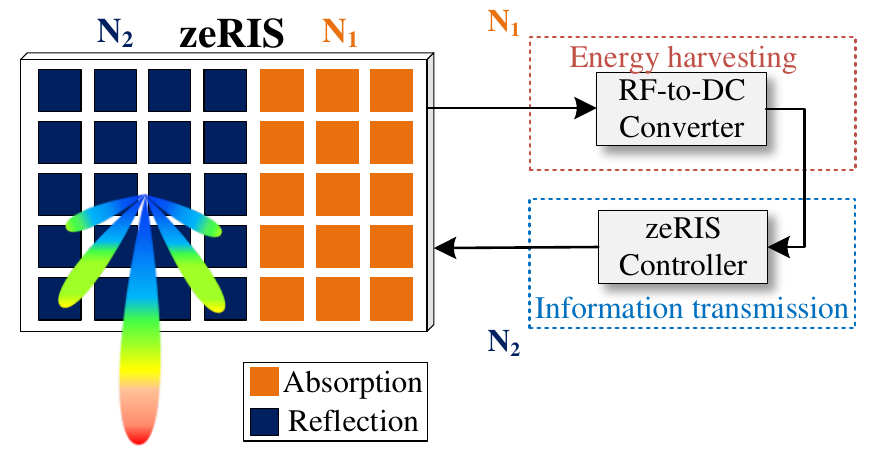}
    \caption{Element splitting (ES)}
    \end{subfigure}
\caption{zeRIS-HaR methods}
\label{Fig.lable}
\end{figure}

\subsubsection{Power splitting}The PS method, as illustrated in Fig. 1a, divides the signal power into two separate streams: the EH stream and the information transmission stream \cite{tegos}. This division is controlled by the tunable PS factor {\small$\rho \in [0, 1]$}. The portion of the received power for EH is {\small$\rho$}, while the information transmission stream's power is proportional to {\small$1-\rho$}.
It should be noted that, since EH and reflection are performed simultaneously in PS, the aforementioned capacitor provides the initial required energy for the zeRIS operation.
As a result, the received signal for the PS case can be expressed as follows
\begin{equation}
\small
\begin{split}
    y_{\mathrm{PS}}= &\sqrt{\ell P_t \left(1-\rho\right) G} \sum_{i=1}^{N} \lvert h_{1i}\rvert \lvert h_{2i}\rvert e^{j \phi_i} x + n,
\end{split}
\end{equation}
where {\small$x$} is the transmitted symbol with unit energy, {\small$P_t$} is the transmit power, {\small$G=G_t G_r$} is the product of the antenna gains, i.e., the transmit and receive antenna, and {\small$N$} is the number of reflecting elements. Furthermore, {\small$n$} describes the additive white Gaussian noise with zero mean and variance equal to {\small$\sigma^2$}, {\small$\ell=\ell_1 \ell_2$} is the end-to-end link's path loss and it is equal to the product of the path losses of the BS-zeRIS and the zeRIS-UE links that are given as
\begin{equation}
\small
    \ell_u = C_0 d_u^{-a_{u}},
\end{equation}
where {\small$u\in \{1,2\}$}, {\small$d_1$} is the BS-zeRIS distance, {\small$d_2$} is the zeRIS-UE distance, and {\small$a_1$}, {\small$a_2$} denote the path loss exponents for the BS-zeRIS and zeRIS-UE channels, respectively. Moreover, {\small$C_0=\frac{\lambda^2}{16 \pi^2}$} is the path loss at the reference distance {\small$d_0$}, where {\small$\lambda$} is the wavelength of the emitted waves, {\small$h_{1i}$} and {\small$h_{2i}$} are the channel coefficients between the BS and the {\small$i$}-th reflecting element and between the {\small$i$}-th reflecting element and the UE, respectively, and  {\small$\phi_i = \omega_i + \mathrm{arg}\left(h_{1i}\right) + \mathrm{arg}\left(h_{2i}\right)$}, where {\small$\omega_i$} is the phase shift induced by the {\small$i$}-th reflecting element, {\small$\mathrm{arg}\left(h_{1i}\right)$} is the phase of {\small$h_{1i}$}, and {\small$\mathrm{arg}\left(h_{2i}\right)$} is the phase of {\small$h_{2i}$}. Therefore, assuming that {\small$\rho$} is equal for all reflecting elements, the instantaneous rate when the PS method is applied and the zeRIS is perfectly configured for the beam-steering functionality, i.e., {\small$\omega_i=-\arg(h_{1i})-\arg(h_{2i})$}, can be expressed as
\begin{equation}\label{RPS}
\small
    R_{\mathrm{PS}} = \mathrm{log_2}\left(1 + \gamma_t G \ell \left(1- \rho \right)  \left| \sum_{i=1}^{N} \lvert h_{1i} \rvert \lvert h_{2i} \rvert \right|^2\right),
\end{equation}
where {\small$\gamma_t=\frac{P_t}{\sigma^2}$} is the transmit SNR.


To effectively utilize a zeRIS, it is critical to determine its power consumption, which results from the configuration of reflecting elements, controller operation, and EH circuitry. Therefore, a key factor that strongly influences the consumption of a zeRIS is the chosen HaR method, as it determines how many reflecting elements are configured for energy absorption and how many contribute to other electromagnetic functionalities. In this direction, when using the PS method, all zeRIS reflecting elements handle both beam-steering and energy absorption. Consequently, the necessary amount of energy that needs to be harvested when the PS method is applied can be expressed as
\begin{equation}\label{EPS}
\small
E_{\mathrm{PS}}=T\left( N P_e + P_{\mathrm{circ}}\right),
\end{equation}
where {\small$T$} represents the time slot duration, {\small$P_e$} denotes the power consumption of each reflecting element, and {\small$P_{\mathrm{circ}}$} denotes the power consumption of the zeRIS controller responsible for setting the induced phase shift for each element \cite{TCOM,venkatesh}. Hence, by employing the linear EH model, the harvested energy when employing the PS technique can be expressed as
\begin{equation}
\small
    Q_{\mathrm{PS}} = T \rho \zeta  P_t G_t l_1 \left| \sum_{i=1}^N \lvert h_{1i} \rvert e^{j \left(\omega_i +\mathrm{arg}\left(h_{1i}\right)\right)} \right| ^2,
\end{equation}
 where {\small$\zeta\in ( 0,1]$} is the energy conversion efficiency. It should be mentioned that {\small$\zeta$} accounts for both the inherent inefficiencies in the EH circuitry and the power losses due to the operational needs of the circuitry. Therefore, {\small$\zeta$} effectively captures the available energy for use after the EH and conversion processes.
 
 As it can be observed, to maximize the amount of harvested energy, the phase shift term {\small$\omega_i$} should be equal to {\small$-\arg(h_{1i})$}, to achieve the maximum channel gain. However, considering that in PS the zeRIS is configured to perform the beam-steering functionality, the channel gain for the harvested energy is equal to {\small$\left| \sum_{i=1}^N \left| h_{1i}\right| e^{j \mathrm{arg}(h_{2i})}\right|$}. It should be noted that, in this work, we have chosen a suboptimal PS scheme, aligning the cascaded BS-RIS-user link, for simplicity and ease of implementation, in contrast to fully-optimized schemes of significant complexity, but also considering that this link is subject to double path loss. Therefore, the amount of harvested energy when PS is applied can be rewritten as
\begin{equation}\label{QPS}
\small
    Q_{\mathrm{PS}} = T \rho \zeta  P_t G_t \ell_1 \left| \sum_{i=1}^N \left| h_{1i}\right| e^{j \mathrm{arg}(h_{2i})}\right|^2.
\end{equation}

\subsubsection{Time switching}
As illustrated in Fig. 1b, in the TS case, the received signal is used solely for EH or receiving information during specific time periods. Specifically, {\small$T$} is divided into two time intervals determined by the splitting factor {\small$\tau \in \left[0,1\right]$}. Within the first time interval, i.e., {\small$[0,\tau T ]$}, the zeRIS is configured for EH, while within the second time interval, i.e., {\small$(\tau T, T ]$}, the RIS is configured for information transmission. Thus, the received signal for TS is given as
\begin{equation}
\small
\begin{split}
y_{\mathrm{TS}} = 
    \begin{cases}
     0,  &  0\leq t \leq \tau T \\
     \sqrt{\ell P_t G} \sum\limits_{i=1}^{N} \lvert h_{1i}\rvert \lvert h_{2i}\rvert e^{j \phi_i } x  +  n,  &  \tau T < t \leq T.
    \end{cases}
\end{split}
\end{equation}
Hence, the instantaneous rate at the receiver when the TS method is applied can be expressed as
\begin{equation}\label{RTS}
\small
R_{\mathrm{TS}} = \left(1-\tau \right) \mathrm{log_2}\left(1 + \gamma_t G \ell   \left| \sum_{i=1}^{N} \lvert h_{1i}\rvert \lvert h_{2i}\rvert  \right|^2\right).
\end{equation}

Accordingly to the PS case, we need to define the energy that needs to be harvested via the TS method. Hence, considering that all reflecting elements will be configured for information transmission only for a specific amount of time, then the required energy for the TS case is given as
\begin{equation}\label{ETS}
\small
    E_{\mathrm{TS}}= T\left(\left(1-\tau\right)N P_e + P_{\mathrm{circ}}\right). 
\end{equation}
Thus, to maximize the absorbed energy within the EH time interval, the phase shift term of each reflecting element is set as {\small$\omega_i = -\arg(h_{1i})$}.  Therefore, in contrast to PS, when TS is applied, the reflecting elements are set in the first time interval to absorb the maximum amount of energy which is given as
\begin{equation}\label{QTS}
\small
    Q_{\mathrm{TS}} = T\left(\tau \zeta P_t G_t \ell_1 \left| \sum_{i=1}^N \left| h_{1i} \right|  \right| ^2\right).
\end{equation}
It should be highlighted that due to the fact that each reflecting element can select a unique value of phase shift to induce, the zeRIS can acquire the maximum amount of energy in the TS method, as it can set the corresponding optimal phase shifts within the two time intervals. However, in PS, by considering that the phase shifts are set for the beam-steering functionality, the maximum amount of energy cannot be harvested, as the reflecting elements are not properly configured.

\subsubsection{Element splitting}
Aside from PS and TS, the large number of reflecting elements enables a zeRIS to harvest energy through the ES method. In more detail, for the ES method, {\small$N_1$} reflecting elements are configured for EH, while the rest of them, i.e., {\small$N_2$} with {\small$N_1+N_2=N$}, are configured for information transmission, as shown in Fig. 1c. Therefore, the received signal for the ES case can be expressed as
\begin{equation}
\small
\begin{split}
    y_{\mathrm{ES}}= \sqrt{\ell P_t G} \sum_{i=N_1+1}^{N} \lvert h_{1i}\rvert  \lvert h_{2i}\rvert  e^{j \phi_i} x + n.
\end{split}
\end{equation}
Thus, the instantaneous rate at the UE when the ES method is applied is given as
\begin{equation}\label{RES}
\small
    R_{\mathrm{ES}} = B\mathrm{log_2}\left(1 + \gamma_t G \ell  \left| \sum_{i=N_1+1}^{N} \lvert h_{1i} \rvert \lvert h_{2i} \rvert \right|^2\right),
\end{equation}
Finally, considering the number of reflecting elements that participate in the beam-steering functionality, the required energy for a zeRIS that performs ES is given as
\begin{equation}\label{EES}
\small
    E_{\mathrm{ES}}= T\left(N_2 P_e + P_{\mathrm{circ}}\right).
\end{equation}
To this end, the harvested energy can be expressed as
\begin{equation}\label{QES}
\small
    Q_{\mathrm{ES}} = T \zeta  P_t G_t \ell_1 \left| \sum_{i=1}^{N_1} \left| h_{1i}\right| \right|^2.
\end{equation}
It should be noted that by dividing the reflecting elements into absorbing and beam-steering elements, the absorbing reflecting elements are configured properly for maximum energy absorption. Finally, it should be highlighted that for the rest of this work, {\small$T$} is assumed to be equal to 1, therefore it is omitted for brevity.

\subsection{zeRIS Deployment}
Considering that the communication performance of a conventional RIS-assisted system is optimized when the RIS is positioned close to the BS or the UE \cite{You}, we investigate two deployment strategies, namely \textit{BS-side} zeRIS or \textit{UE-side} zeRIS. In more detail, in the \textit{BS-side} case, the BS chooses a nearby zeRIS with which it shares a line-of-sight (LoS) channel, i.e., {\small$\lvert h_{1i} \rvert=1$} and {\small$\arg(h_{1i})= \frac{2 \pi d_1}{\lambda}$}, while the zeRIS-UE channel {\small$\lvert h_{2i} \rvert$} is assumed to be a random variable (RV) following the Nakagami-{\small$m$} distribution with shape parameter {\small$m$} and scale parameter {\small$\Omega$} . Accordingly, for the \textit{UE-side} case, the zeRIS is deployed near the UE with which it shares a LoS link, i.e., {\small$\lvert h_{2i} \rvert=1$} and {\small$\arg(h_{2i})= \frac{2 \pi d_2}{\lambda}$}, while the BS-zeRIS channel {\small$\lvert h_{1i} \rvert$} is assumed to be an RV following the Nakagami-{\small$m$} distribution with shape parameter {\small$m$} and scale parameter {\small$\Omega$}. Thus, the moments of {\small$\lvert h_{2i} \rvert$} in the \textit{BS-side} and {\small$\lvert h_{1i} \rvert$} in the \textit{UE-side} can be expressed as {\small$\mathbb{E}[\left| h_{ui}\right|^n]= \dfrac{\Gamma(m+\frac{n}{2})}{\Gamma(m)} \left(\frac{\Omega}{m}\right)^{\frac{n}{2}}$}. It should be highlighted, that our objective in detailing both \textit{BS-side} and \textit{UE-side} zeRIS scenarios is not a mere comparison, but to provide a tailored framework rooted in the PWE concept, where the decision on which zeRIS should be used often depends on the specific application and, crucially, on user reachability.

\section{Theoretical Analysis}
A zeRIS must acquire a sufficient amount of energy to perform different functionalities such as beam-steering. Therefore, in order to conduct a comprehensive assessment of the reliability of a communication system that relies on a zeRIS, we utilize a performance metric termed as \textit{joint energy-data rate outage probability}. This metric takes into account both the energy outage event, which describes that the zeRIS has failed to harvest sufficient energy for its operation, and the data rate outage event, which indicates that the rate falls below a predetermined rate threshold {\small$R_{\mathrm{thr}}$} in bit/s/Hz. Furthermore, considering that different EH techniques can be employed by the zeRIS to obtain the necessary energy, we also calculate the network's energy efficiency for each examined HaR method and zeRIS deployment location (i.e., \textit{BS-side} or \textit{UE-side}). Thus, in this section, we provide analytical expressions for the joint energy-data rate outage probability of the evaluated HaR methods, as well as the network's energy efficiency for the cases where the zeRIS is deployed either at the \textit{BS-side} or at the \textit{UE-side}.

\begin{definition}
The joint energy-data rate outage probability of a zeRIS-assisted system is defined as the union of the energy outage event, i.e., the zeRIS has not harvested the required amount of energy for its operation, and the data rate outage event, i.e., the received SNR is lower than a predefined SNR threshold \cite{tegos}. Therefore, we present analytic expressions for the joint energy-data rate outage probability which can be expressed as 
\begin{equation}\label{PGEN}
\small
    P_{q_1}^{q_2} = \Pr \left(Q_{q_2} \leq E_{q_2} \ \cup \ R_{q_2} \leq R_{\mathrm{thr}}\right),
\end{equation}
where {\small$Q_{q_2}$} is the amount of harvested energy, {\small$E_{q_2}$} is the required amount of energy for the zeRIS operation, {\small$R_{q_2}$} is the instantaneous rate at the UE, {\small$q_1 \in \{\mathrm{B},\mathrm{U}\}$} indicates if the zeRIS is deployed \textit{BS-side} or \textit{UE-side}, respectively, and {\small$q_2 \in \{\mathrm{PS, TS, ES}\}$} indicates which HaR method is applied. 
\end{definition}

In addition, the escalating demand for wireless communication necessitates the optimization of energy consumption in networks, particularly in the context of zeRISs, in order to minimize their environmental impact. Thus, it is imperative to quantify the required transmit power {\small$P_t$} to achieve a specific outage performance. 
\begin{definition}
The energy efficiency of a zeRIS-assisted network is defined as the ratio of the target rate multiplied by the complementary probability of the joint energy-data rate outage probability to the transmit power {\small$P_t$} and can be expressed as
\begin{equation}
\small
    \mathcal{E}_{q_1}^{q_2}= \dfrac{R_{\mathrm{thr}}}{P_t} \left(1- P_{q_1}^{q_2}\right) \quad \left[\text{bit}/\text{J}/ \text{Hz}\right].
\end{equation}
\end{definition}

\begin{remark}
The inclusion of the joint energy-data rate outage probability within the definition of energy efficiency enables the quantitative assessment of the performance of a zeRIS-assisted network with regard to both communication performance and energy consumption.
\end{remark}

\subsection{BS-side zeRIS}
For the case where the zeRIS is placed near to the BS, the joint energy-data rate outage probability for the examined EH techniques, is given in the following propositions. First, the PS case is presented.
\begin{proposition} \label{prop:1}
The joint energy-data rate outage probability for a \textit{BS-side} zeRIS that applies the PS technique can be approximated as
\begin{equation}
\small
\begin{split}
    {P^{\mathrm{PS}}_{\mathrm{B}}} &\approx \frac{1}{2} \left[\mathrm{erf}\left(\frac{w_1 +\mu}{\sqrt{2 \sigma^2}} \right) +  \mathrm{erf}\left(\frac{w_1 -\mu}{\sqrt{2 \sigma^2}} \right)\right] + \frac{\gamma\left(k_{\mathrm{PS}},\frac{w_2}{\theta_{\mathrm{PS}}}\right)}{\Gamma(k_{\mathrm{PS}})} \\
    \\& -\frac{\gamma\left(k_{\mathrm{PS}},\frac{w_2}{\theta_{\mathrm{PS}}}\right)}{2\Gamma(k_{\mathrm{PS}})} \left[\mathrm{erf}\left(\frac{w_1 +\mu}{\sqrt{2 \sigma^2}} \right) +  \mathrm{erf}\left(\frac{w_1 -\mu}{\sqrt{2 \sigma^2}} \right)\right]
    \end{split}
\end{equation}
with
\begin{equation}
\small
k_{\mathrm{PS}}= \dfrac{N \left(\frac{\Gamma(m+\frac{1}{2})}{m}\right)^2}{m - \left(\frac{\Gamma(m+\frac{1}{2})}{m}\right)^2}, 
\end{equation}
\begin{equation}
\small
\theta_{\mathrm{PS}}=\dfrac{\sqrt{\Omega \Gamma^2(m)} \left(\sqrt{m} - \frac{1}{\sqrt{m}}\left(\frac{\Gamma(m+\frac{1}{2})}{\Gamma(m)}\right)^2 \right)}{\Gamma(m+\frac{1}{2})},
\end{equation}
where {\small$\mathrm{erf}(\cdot)$} is the error function.
Moreover,
\begin{equation}
\small
\mu = N \left(\frac{I_1(\kappa)}{I_0(\kappa)\left(K+1\right)} + \frac{K}{K+1} \right),
\end{equation}
\begin{equation}
\small
\sigma^2= \frac{N}{2} \left(\frac{I_2(\kappa)}{I_0(\kappa)\left(K+1\right)} + \frac{K}{K+1} \right),
\end{equation}
\begin{equation}
\small
w_1=\sqrt{\dfrac{N P_e+P_{\mathrm{circ}}}{\rho \zeta P_t G_t l_1}} ,
\end{equation}
\begin{equation}
\small
w_2= \sqrt{\dfrac{2^{R_{\mathrm{thr}}}-1}{\gamma_t G l (1-\rho)}},
\end{equation}
where {\small$K\approx \frac{\sqrt{m^2 - m}}{m - \sqrt{m^2-m}}$} is the Rice factor, and {\small$\kappa\in \left[0, \infty\right)$} is the von Mises concentration parameter. 
\end{proposition}

\begin{IEEEproof}
The proof is found in Appendix A.
\end{IEEEproof}

As it can be observed, the system performance for the \textit{BS-side PS} case strongly depends on whether the communication link has a direct connection or is affected by obstacles. Notably, even when the zeRIS-UE link is used only for communication, the conditions of this link still significantly affect the absorption functionality. This emphasizes the importance of the channel conditions and points out unique considerations in zeRIS-assisted communications.

In contrast to the PS method, where the zeRIS simultaneously performs the beam-steering and the absorption functionality, in the TS method, the zeRIS is configured for energy absorption until it harvests the required amount of energy, and then all its elements are reconfigured to serve the data transmission via beam-steering. Thus, in the following proposition, we provide the joint energy-data rate outage probability for the case where the TS method is applied from a \textit{BS-side} zeRIS.
\begin{proposition}
The joint energy-data rate outage probability for a \textit{BS-side} zeRIS that applies the TS method can be approximated as
\begin{equation} \label{OPBTS}
\small
\begin{split}
{P^{\mathrm{TS}}_{\mathrm{B}}} \approx 
    \begin{cases}
     1, & N < w_3  \\
    \dfrac{\gamma(k_{\mathrm{TS}},\frac{w_4}{\theta_{\mathrm{TS}}} )}{\Gamma(k_{\mathrm{TS}})},   &  \text{otherwise},
    \end{cases}
\end{split}
\end{equation}
where {\small$k_{\mathrm{TS}}=k_{\mathrm{PS}}$}, {\small$\theta_{\mathrm{TS}}=\theta_{\mathrm{PS}}$}, {\small$w_3=\sqrt{\dfrac{\left(1-\tau\right)NP_e + P_{\mathrm{circ}}}{\tau \zeta P_t G_t \ell_1} }$} and {\small$w_4=\sqrt{\dfrac{2^{\frac{R_{\mathrm{thr}}}{1-\tau}} -1}{\gamma_t G \ell}}$}.
\end{proposition}
\begin{IEEEproof}
    By utilizing \eqref{RTS}, \eqref{ETS}, \eqref{QTS}, and {\small$\lvert h_{1i}\rvert=1$}, the joint energy-data rate outage probability of a network equipped with a \textit{BS-side} zeRIS that applies TS can be expressed as
\begin{equation}\label{PJTS1}
\small
\begin{split}
    {P^{\mathrm{TS}}_{\mathrm{B}}} & = \Pr \Bigg( \tau \zeta P_t G_t \ell_1 N^2 \leq \left(1-\tau\right)N P_e + P_{\mathrm{circ}} \\
    &\cup \ \left(1-\tau \right) \mathrm{log_2}\Bigg(1 + \gamma_t G \ell  \left| \sum_{i=1}^{N} \lvert h_{2i} \rvert \right|^2\Bigg) \leq R_{\mathrm{thr}} \Bigg) .
\end{split}
\end{equation}
As it can be observed, the first inequality does not include any RV, thus the probability can be rewritten as
\begin{equation}
\small
\begin{split}
{P^{\mathrm{TS}}_{\mathrm{B}}} = 
    \begin{cases}
     1,  &  N < w_3 \\
    \Pr\left(\sum_{i=1}^{N} \lvert h_{2i} \rvert \leq w_4 \right), &  \text{otherwise}.
    \end{cases}
\end{split}
\end{equation}
Therefore, by invoking the moment-matching method in a similar way as in \textit{Proposition 1}, the probability in the second branch can be derived, which concludes the proof.
\end{IEEEproof}

\begin{remark}
    Considering \eqref{OPBTS} which shows monotonicity when $N \geq w_3$ and coincides with the CDF of the Gamma distribution, an increase in $\tau$ leads to a corresponding increase in $w_4$. This implies that the joint energy-data rate outage probability will continuously increase. To this end, the optimal time splitting factor $\tau_{\mathrm{B}}^{*}$ can be derived by solving the equation $N=w_3$. Thus, after some algebraic manipulations, we can obtain the optimal time splitting factor $\tau_{\mathrm{B}}^{*}$ which is equal to
\begin{equation}
\small
    \tau_\mathrm{B}^{*}= \frac{NP_e + P_{\mathrm{circ}}}{NP_e + N^2 \zeta P_t G_t l_1}.
\end{equation}
\end{remark}

In addition to the PS and the TS methods, a zeRIS can also operate by dividing its reflecting elements into two parts: {\small$N_1$} reflecting elements performing the absorption functionality to assist in the EH, meaning that their induced phase shift is set as {\small$\omega_i= -\arg(h_{1i})$}, and {\small$N_2$} reflecting elements performing the beam-steering functionality for information transmission, meaning that their induced phase shift is set as {\small$\omega_i= -\arg(h_{1i})-\arg(h_{2i}) $}. To this end, in the following proposition, we derive the joint energy-data rate outage probability for the case where a \textit{BS-side} zeRIS harvests energy through the ES method.
\begin{proposition}
The joint energy-data rate outage probability for a \textit{BS-side} zeRIS that applies the ES technique can be approximated as
\begin{equation}\label{opbes}
\small
\begin{split}
{P^{\mathrm{ES}}_{\mathrm{B}}} \approx
    \begin{cases}
     1,  &  N_1 < w_5 \\
    \dfrac{\gamma\left(m_{\mathrm{es}}(2,N_2),\frac{m_{\mathrm{es}}(2,N_2) w_6^2}{\Omega_{\mathrm{es}}(N_2)} \right)}{\Gamma\left(m_{\mathrm{es}}(2,N_2)\right)},  & \text{otherwise},
    \end{cases}
\end{split}
\end{equation}
where  
\begin{equation}\label{mts}
\small
    m_{\mathrm{es}}(u,N_i) =  \frac{(\Omega_{\mathrm{es}}(N_i))^2}{f(u,N_{i})-(\Omega_{\mathrm{es}}(N_i))^2},
\end{equation}
and
\begin{equation}\label{omegaes}
\small
    \Omega_{\mathrm{es}}(N_i) = N_{i}\left(\mathbb{E}[\left| h_{ui}\right|^2] +(N_{i}-1)\mathbb{E}^2[\left| h_{ui}\right|]\right).
\end{equation}
Furthermore,  {\small$w_5=\sqrt{\dfrac{N_2 P_e + P_{\mathrm{circ}}}{\zeta P_t G_t \ell_1} }$}, {\small$w_6=\sqrt{\dfrac{2^{R_{\mathrm{thr}}}-1}{\gamma_t G \ell}}$}, and {\small$ f(u,N_{i})$} is given in \eqref{f} at the top of the next page.
\end{proposition}

\begin{IEEEproof}
    By substituting \eqref{RES}, \eqref{EES}, and \eqref{QES} in \eqref{PGEN}, we obtain 
    \setcounter{equation}{31}
\begin{equation}\label{PJES1}
\small
\begin{split}
    {P^{\mathrm{ES}}_{\mathrm{B}}} &= \Pr \Bigg( \zeta P_t G_t \ell_1 N_1^2 \leq N_2 P_e + P_{\mathrm{circ}} \\
    &  \cup \ \mathrm{log_2}\left(1 + \gamma_t G \ell  \left| \sum_{i=N_1+1}^{N} \lvert h_{2i} \rvert \right|^2\right) \leq R_{\mathrm{thr}} \Bigg).
\end{split}
\end{equation}
Considering that {\small$N_2$} is not necessarily large, we cannot guarantee that the moment-matching technique can offer a tight approximation for {\small${P^{\mathrm{ES}}_{\mathrm{B}}}$}. Therefore, by utilizing the results provided in \cite{Yacoub} the sum of {\small$N_i$} independent and identically distributed Nakagami-{\small$m$} RVs can be approximated by a Nakagami-{\small$m$} RV with shape parameter {\small$m_{\mathrm{es}}(u,N_i)$} and scale parameter {\small$\Omega_{\mathrm{es}}(N_i)$}, respectively. Finally, considering the cumulative density function (CDF) of the Nakagami-{\small$m$} distribution which is equal to
\begin{equation}
\small
    F_n(x) = \frac{\gamma\left(m_{\mathrm{es}}(u,N_i),\frac{m_{\mathrm{es}}(u,N_i) x^2}{\Omega_{\mathrm{es}}(N_i)} \right)}{\Gamma\left(m_{\mathrm{es}}(u,N_i)\right)},
\end{equation} 
{\small${P^{\mathrm{ES}}_{\mathrm{B}}}$} can be obtained, which concludes the proof.
\end{IEEEproof}

\begin{figure*}[t]
\begin{equation}\label{f} 
\small
\tag{31}
\begin{split}
    f(u,N_i) & =  N_{i}\left( \mathbb{E}[{\left| h_{ui}\right|}^4]+ 4(N_{i}-1)\mathbb{E}[{\left| h_{ui}\right|}^3]\mathbb{E}[{\left| h_{ui}\right|}]+3(N_{i}-1)\mathbb{E}^2[{\left| h_{ui}\right|}^2]+ 6(N_{i}-1)(N_{i}-2)\mathbb{E}[{\left| h_{ui}\right|}^2]\mathbb{E}^2[{\left| h_{ui}\right|}] \right.  \\
    & \qquad \qquad \left. +(N_{i}-1)(N_{i}-2)(N_{i}-3)\mathbb{E}^4[{\left| h_{ui}\right|}] \vphantom{\mathbb{E}[{\left| h_{ui}\right|}^4]} \right)
\end{split}
\end{equation}
\hrule
\end{figure*}

\begin{remark}
    Considering the monotonicity of \eqref{opbes} for the case where $N_1\geq w_5$, by increasing $N_1$, $N_2$ decreases, which means that the joint energy-data rate outage probability will tend to 1. To this end, the optimal number of reflecting elements $N_1^{*}$ can be derived by solving the equation $N_1=w_5$. Thus, by replacing $N_2$ with $N-N_1$, after some algebraic manipulations, we can obtain the optimal number of absorbing reflecting elements $N_{1,\mathrm{B}}^{*}$, which can be expressed as
\begin{equation}
\small
N_{1,\mathrm{B}}^{*} = \frac{-P_e + \sqrt{P_e^2 + 4\zeta P_t G_t l_1 \left( N P_e + P_{\mathrm{circ}}\right)}}{2 \zeta P_t G_t l_1}.
\end{equation}
\end{remark}

\subsection{UE-side zeRIS} As stated in the literature, due to the double path loss model, as shown in (2), a zeRIS can also be deployed in the proximity of the UE to establish LoS zeRIS-UE links, i.e., {\small$\lvert h_{2i} \rvert=1$},  and offer reliable communication \cite{You}. Hence, by assuming that {\small$\lvert h_{1i} \rvert$} is a Nakagami-{\small$m$} RV with shape parameter {\small$m$}, and scale parameter {\small$\Omega$}, then the joint energy-data rate outage probability for all the examined EH techniques is given in the following proposition.
\begin{proposition}
The joint energy-data rate outage probability for the PS protocol can be approximated as
    \begin{equation}
  \small
        P_{\mathrm{U}}^{\mathrm{PS}} \approx \frac{1}{\Gamma(k_{\mathrm{PS}})} \gamma\left(k_{\mathrm{PS}},\frac{\mathrm{max}\left(w_{1},w_2\right)}{\theta_{\mathrm{PS}}}\right).
    \end{equation}
\end{proposition}
\begin{IEEEproof} 
By substituting \eqref{RPS}, \eqref{QPS}, \eqref{EPS} in \eqref{PGEN}, and by taking into account that {\small$\lvert h_{2i} \rvert=1$}, the joint energy-data rate outage probability for a \textit{UE-side zeRIS}-assisted network that applies the PS method can be expressed as
\begin{equation}
\small
\begin{split}
P_{\mathrm{U}}^{\mathrm{PS}} & = \Pr \Bigg(\rho \zeta  P_t G_t \ell_1 \left| \sum_{i=1}^N \lvert h_{1i} \rvert  \right| ^2 \leq N P_n + P_{\mathrm{circ}} \\
&\cup \ \mathrm{log_2}\Bigg(1 + \gamma_t G \ell \left(1- \rho \right) \left| \sum_{i=1}^{N} \lvert h_{1i} \rvert \right|^2\Bigg) \leq R_{\mathrm{thr}}\Bigg),
\end{split}
\end{equation}
which after some algebraic manipulations can be rewritten as
\begin{equation}
\small
\begin{split}
P_{\mathrm{U}}^{\mathrm{PS}} & = \Pr \Bigg( \sum_{i=1}^N \left| h_{1i} \right| \leq w_1  \cup \  \sum_{i=1}^N \left| h_{1i}\right| \leq w_2\Bigg).
\end{split}
\end{equation}
It can be observed that {\small$Z_2=\sum_{i=1}^N \left| h_{1i}\right|$} is upper bounded in both events. Specifically, the union of these events occurs when {\small$Z_2$} is lower than the maximum of these upper bounds. Hence, {\small$P_{\mathrm{B}}^{\mathrm{PS}}$} can be further expressed as
\begin{equation}
\small
P_{\mathrm{U}}^{\mathrm{PS}} =  \Pr \left(Z_2 \leq \mathrm{max}\left(w_1,w_2\right) \right).
\end{equation}
Thus, by invoking the moment-matching technique as shown in \textit{Proposition 1}, {\small$Z_2$} can be tightly approximated by a gamma-distributed RV with shape parameter {\small$k_{\mathrm{PS}}$} and scale parameter {\small$\theta_{\mathrm{PS}}$}, respectively, and, thus, {\small$P_{\mathrm{U}}^{\mathrm{PS}}$} can be calculated through \eqref{CDF}, which concludes the proof.
\end{IEEEproof}
\begin{remark}
By setting {\small$w_1=w_2$}, we can obtain the optimal power-splitting factor that minimizes the joint energy-data rate outage probability for the \textit{UE-side} PS case, which is equal to
\begin{equation}\label{rho_opt}
\small
    \rho_\mathrm{U}^{*} = \frac{1}{1 + \frac{\sigma^2 \zeta \left(2^{R_{\mathrm{thr}}} - 1 \right)}{\left(N P_e + P_{\mathrm{circ}}\right)G_r l_2}}.
\end{equation}
\end{remark}
As it can be observed, \eqref{rho_opt} highlights the trade-off between EH and information decoding. To this end, adjusting$\rho^{*}$allows for the optimization of both EH efficiency and data rate, depending on the energy demand and supply.

Next, we provide the joint energy-data rate outage probability for the case where a \textit{UE-side} zeRIS harvests energy through the TS method.
\begin{proposition}
The joint energy-data rate outage probability when the TS method is applied can be approximated as
\begin{equation}\label{UTS}
\small
        P_{\mathrm{U}}^{\mathrm{TS}} \approx \frac{1}{\Gamma(k_{\mathrm{TS}})} \gamma\left( k_{\mathrm{TS}},\frac{\mathrm{max}\left(w_{3},w_4\right)}{\theta_{\mathrm{TS}}} \right).
    \end{equation}
\end{proposition}
\begin{IEEEproof}
Following a similar procedure with \textit{Proposition 4}, by substituting \eqref{RTS}, \eqref{ETS}, and \eqref{QTS} in \eqref{PGEN}, and after some algebraic manipulations, the joint energy-data rate outage probability can be written as
\begin{equation}
\small
P_{\mathrm{U}}^{\mathrm{TS}} = \Pr \left(Z_2 \leq w_{3} \ \cup \ Z_2 \leq w_4 \right).
\end{equation}
By using the moment-matching technique as presented in \textit{Proposition 1}, the joint energy-data rate outage probability can be derived as in \eqref{UTS}, which concludes the proof.
\end{IEEEproof}
\begin{remark}
Similarly with the \textit{UE-side} PS method, by setting {\small$w_3$} and {\small$w_4$} equal, the joint-energy-data rate outage probability can be minimized. Therefore, by following some algebraic manipulations, we arrive at 
\begin{equation}
\small
    \tau_{\mathrm{U}}^{*} - \frac{\zeta \sigma^2 + \left(N P_e + P_{\mathrm{circ}}\right)G_r l_2}{N P_e + 2^{\frac{R_{\mathrm{thr}}}{1-\tau_{\mathrm{U}}^{*}}} \zeta \sigma^2 } = 0,
\end{equation}
from which the optimal splitting factor $\tau_{\mathrm{U}}^{*}$ can be obtained numerically, minimizing the joint energy-data rate outage probability for the UE-side TS case. This equation further underscores the trade-off between EH and information decoding.
\end{remark}

Finally, we derive the joint energy-data rate outage probability for the case where a \textit{UE-side} zeRIS applies the ES method for EH.

\begin{proposition}
The joint energy-data rate outage probability when the ES method is applied can be approximated by \eqref{UES} at the top of the next page.
\end{proposition}
\begin{IEEEproof}
    Similarly with the PS and TS methods, by substituting \eqref{RES}, \eqref{EES}, and \eqref{QES} in \eqref{PGEN}, the joint energy-data rate outage probability can be expressed as
    \setcounter{equation}{43}
\begin{equation}\label{PJES_1}
\small
\begin{split}
P_{\mathrm{U}}^{\mathrm{ES}} = \Pr \left( \sum_{i=1}^{N_1} \left| h_{1i}\right| \leq w_5 \ \cup \ \sum_{j=N_1+1}^{N} \left| h_{1j}\right| 
\leq w_6\right) .
\end{split}
\end{equation}
As it can be observed, the RVs that are upper bounded in \eqref{PJES_1} are different and independent of each other, due to the fact that different reflecting elements perform the absorption and the beam-steering functionality. Therefore, the above probability can be rewritten as
\begin{equation}
\small
\begin{split}
P_{\mathrm{U}}^{\mathrm{ES}} & = \Pr \left( \sum_{i=1}^{N_1} \left| h_{1i}\right| \leq w_5\right) +\Pr \left(\sum_{j=N_1+1}^{N} \left| h_{1j}\right| \leq w_6\right) \\
&  - \Pr \left(\sum_{i=1}^{N_1} \left| h_{1i}\right| \leq w_5\right) \Pr \left(\sum_{j=N_1+1}^{N} \left| h_{1j}\right| \leq 
w_6\right).
\end{split}
\end{equation}
Again, considering that {\small$N_1$} and {\small$N_2$} are not necessarily large, by following the same procedure as in \textit{Proposition 3}, \eqref{UES} can be derived, which concludes the proof.
\end{IEEEproof}

\begin{figure*}[t!]
\begin{equation}
\small
\tag{43}
\begin{split}\label{UES}
    P_{\mathrm{U}}^{\mathrm{ES}} \approx \frac{\gamma\left(m_{\mathrm{es}}(1,N_1),\frac{m_{\mathrm{es}}(1,N_1)}{ \Omega_{\mathrm{es}}(N_1)} (w_5)^2\right)}{\Gamma(m_{\mathrm{es}}(1,N_1))} &+\frac{\gamma\left(m_{\mathrm{es}}(1,N_2),\frac{m_{\mathrm{es}}(1,N_2)}{ \Omega_{\mathrm{es}}(N_2)} \left( w_6 \right)^2\right)}{\Gamma(m_{\mathrm{es}}(1,N_2))} \\
    &- \left( \frac{\gamma\left(m_{\mathrm{es}}(1,N_1),\frac{m_{\mathrm{es}}(1,N_1)}{ \Omega_{\mathrm{es}}(N_1) } (w_5)^2\right)}{\Gamma(m_{\mathrm{es}}(1,N_1))}\right)\left( \frac{\gamma\left(m_{\mathrm{es}}(1,N_2),\frac{m_{\mathrm{es}}(1,N_2)}{ \Omega_{\mathrm{es}}(N_2)} \left( w_6 \right)^2\right)}{\Gamma(m_{\mathrm{es}}(1,N_2))}\right)
\end{split}
\end{equation}
\hrule
\end{figure*}
Similar to the \textit{BS-side} ES case, \eqref{UES} takes into account that different zeRIS reflecting elements have distinct roles in absorption and data transmission, which significantly affects the joint outage probability. Hence, (41) emphasizes the critical importance of optimizing zeRIS element allocation to simultaneously improve both energy efficiency and data rate performance in the network.

\begin{table}[!ht]
	\renewcommand{\arraystretch}{1.00}
	\caption{\textsc{ Simulation Results Parameters}}
	\label{values_sim}
	\centering
	\begin{tabular}{lll}
		\hline
		\bfseries Parameter & \bfseries Notation & \bfseries Value \\
		\hline\hline
		Path loss @ reference distance	          &  {\small$C_0$}  	    & {\small$-30$} dB 		    						\\
		Energy conversion efficiency			  &  {\small$\zeta$}			& {\small$0.65$} 									\\
		Element consumption			  &  {\small$P_e$}			& {\small$2$} {\small$\mu$}W 									\\
             Controller consumption            &  {\small$P_{\mathrm{circ}}$}			& {\small$50$} mW 	                                            \\
		Reference distance				  &  {\small$d_0$}		& {\small$1$} m									\\
		Noise variance				  &  {\small$\sigma^2$}		& {\small$-100$} dB 						\\
        BS Antenna gain   &  {\small$G_{t}$}    & {\small$4$} dB 		    						\\
		UE Antenna gain &  {\small$G_r$} 		& {\small$0$} dB 		    						\\
		Path loss exponent (with fading)	  &  {\small$a_u$}		& {\small$2.5$}		  	  						\\
		Shape parameter			  &  {\small$m$}			& {\small$2$} 						\\
		Spread parameter			  &  {\small$\Omega$}			& {\small$1$}  						\\
		Concentration parameter	  &  {\small$\kappa$}    		& {\small$3$}	\\
		\hline
	\end{tabular}
\end{table}

\section{Numerical Results}
In this section, we evaluate the performance of a downlink zeRIS-assisted communication scenario in terms of joint energy-data rate outage probability and energy efficiency. Specifically, both the \textit{BS-side} and \textit{UE-side} zeRIS cases are investigated in terms of joint energy-data rate outage performance, and the most appropriate HaR method to be applied in both cases is determined in terms of energy efficiency. In order to derive the numerical results, we set the parameters of the analyzed system model as shown in Table \ref{values_sim}. It should be highlighted that {\small$\kappa$} is set to be 3 in order to describe a wireless propagation environment with non-isotropically distributed scatterers that impact the phase of the received signal \cite{Abdi}. Furthermore, unless otherwise stated, the transmit power {\small$P_t$} is set at {\small$0.5$} W, while the channel affected by small-scale fading in both \textit{BS-side} and \textit{UE-side} scenarios is assumed to follow Nakagami-{\small$m$} distribution with shape parameter {\small$m=2$} and scale parameter {\small$\Omega=1$}. In addition, the path loss exponent of the channel affected by small-scale fading in the aforementioned scenarios is set equal to {\small$2.5$}, while the values of the reflecting elements' consumption and the zeRIS controller consumption are set as in \cite{TCOM}.  Finally, we employ Monte Carlo simulations to verify the accuracy of the derived analytical results, where the simulation results are illustrated as marks, whereas the analytical results are illustrated as solid and dashed lines.

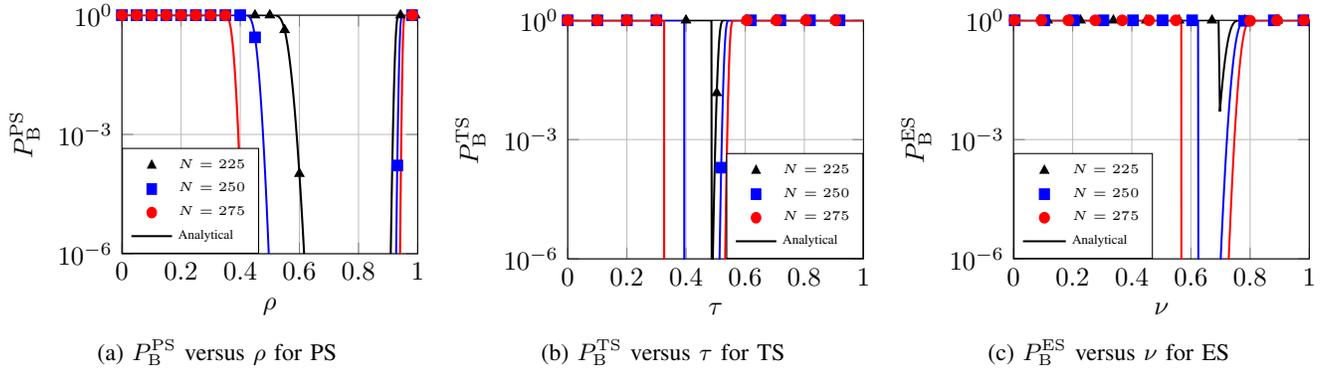
\begin{figure*}
\vspace{-0.2cm}
\begin{subfigure}[t]{.32\textwidth}
	\centering
	\begin{tikzpicture}
	\begin{semilogyaxis}[
	width=0.95\linewidth,
	xlabel = {$\rho$},
	ylabel = {${P^{\mathrm{PS}}_{\mathrm{B}}}$},
	xmin = 0,
	xmax = 1,
	ymin = 0.000001,
	ymax = 1,
	grid = major,
	legend image post style={xscale=0.9},
	legend cell align = {left},
      legend style={at={(0,0)},anchor=south west, font = \tiny},
	]
	\addplot[
	black,
        mark=triangle*,
	mark repeat = 50,
	mark size = 2,
        only marks,
	]
	table {data/PS_BS_225.dat};
	\addlegendentry{$N=225$}
	\addplot[
	blue,
        mark=square*,
	mark repeat = 50,
	mark size = 2,
        only marks,
	]
	table {data/PS_BS_250.dat};
	\addlegendentry{$N=250$}

	\addplot[
	red,
        mark=*,
	mark repeat = 50,
	mark size = 2,
        only marks,
	]
	table {data/PS_BS_275.dat};
	\addlegendentry{$N=275$}
 	\addplot[
	black,
        no marks,
        line width = 0.775pt,
	style = solid,
	]
	table {data/PS_BS_225.dat};
 	\addlegendentry{Analytical}
 	\addplot[
	blue,
        no marks,
        line width = 0.75pt,
	style = solid,
	]
	table {data/PS_BS_250.dat};

  	\addplot[
	red,
        no marks,
        line width = 0.8pt,
	style = solid,
	]
	table {data/PS_BS_275.dat};
	\end{semilogyaxis}
	\end{tikzpicture}
        \caption{ ${P^{\mathrm{PS}}_{\mathrm{B}}}$ versus $\rho$  for PS}
	\label{fig:PBPS}
\end{subfigure}
\begin{subfigure}[t]{.32\textwidth}
	\centering
	\begin{tikzpicture}
	\begin{semilogyaxis}[
	width=0.95\linewidth,
	xlabel = {$\tau$},
	ylabel = {${P^{\mathrm{TS}}_{\mathrm{B}}}$},
	xmin = 0,
	xmax = 1,
	ymin = 0.000001,
	ymax = 1,
	grid = major,
	legend image post style={xscale=0.9},
	legend cell align = {left},
      legend style={at={(1,0)},anchor=south east, font = \tiny},
	]
 
	\addplot[
	black,
        mark=triangle*,
	mark repeat = 100,
	mark size = 2,
        only marks,
	]
	table {data/TS_BS_225.dat};
	\addlegendentry{$N=225$}
	\addplot[
	blue,
        mark=square*,
	mark repeat = 100,
	mark size = 2,
        only marks,
	]
	table {data/TS_BS_250.dat};
	\addlegendentry{$N=250$}

	\addplot[
	red,
        mark=*,
	mark repeat = 100,
	mark size = 2,
        only marks,
	]
	table {data/TS_BS_275.dat};
	\addlegendentry{$N=275$}
 	\addplot[
	black,
        no marks,
        line width = 0.775pt,
	style = solid,
	]
	table {data/TS_BS_225.dat};
 	\addlegendentry{Analytical}

 	\addplot[
	blue,
        no marks,
        line width = 0.75pt,
	style = solid,
	]
	table {data/TS_BS_250.dat};

  	\addplot[
	red,
        no marks,
        line width = 0.8pt,
	style = solid,
	]
	table {data/TS_BS_275.dat};
	\end{semilogyaxis}
	\end{tikzpicture}
         \caption{ ${P^{\mathrm{TS}}_{\mathrm{B}}}$ versus $\tau$ for TS}
	\label{fig:PBΤS}
\end{subfigure}
\begin{subfigure}[t]{.32\textwidth}
	\centering
	\begin{tikzpicture}
	\begin{semilogyaxis}[
	width=0.95\linewidth,
	xlabel = {$\nu$},
	ylabel = {${P^{\mathrm{ES}}_{\mathrm{B}}}$},
	xmin = 0,
	xmax = 1,
	ymin = 0.000001,
	ymax = 1,
	grid = major,
	legend image post style={xscale=0.95},
	legend cell align = {left},
      legend style={at={(0,0)},anchor=south west, font = \tiny},
	]
	\addplot[
	black,
        mark=triangle*,
	mark repeat = 25,
	mark size = 2,
        only marks,
	]
	table {data/ES_BS_225.dat};
	\addlegendentry{$N=225$}
	\addplot[
	blue,
        mark=square*,
	mark repeat = 25,
	mark size = 2,
        only marks,
	]
	table {data/ES_BS_250.dat};
	\addlegendentry{$N=250$}

	\addplot[
	red,
        mark=*,
	mark repeat = 25,
	mark size = 2,
        only marks,
	]
	table {data/ES_BS_275.dat};
	\addlegendentry{$N=275$}

 	\addplot[
	black,
        no marks,
        line width = 0.75pt,
	style = solid,
	]
	table {data/ES_BS_225.dat};
 	\addlegendentry{Analytical}

 	\addplot[
	blue,
        no marks,
        line width = 0.75pt,
	style = solid,
	]
	table {data/ES_BS_250.dat};

  	\addplot[
	red,
        no marks,
        line width = 0.75pt,
	style = solid,
	]
	table {data/ES_BS_275.dat};
	\end{semilogyaxis}
	\end{tikzpicture}
        \caption{ ${P^{\mathrm{ES}}_{\mathrm{B}}}$ versus $\nu$  for ES}
	\label{fig:ES}
\end{subfigure}
\caption{Joint energy-data rate outage probability for \textit{BS-side} zeRIS with {\small$R_{\mathrm{thr}}$}={\small$3.46 $} bit/s/Hz}
\end{figure*}

Fig. \ref{fig:PBPS} illustrates the performance of a \textit{BS-side} zeRIS-assisted network for the case where the zeRIS harvests energy through the PS method. In more detail, the joint energy-data rate outage probability is illustrated versus the PS factor {\small$\rho$} for different values of {\small$N$}, representing the number of reflecting elements in the zeRIS. It should be mentioned that for the \textit{BS-side} zeRIS case, it is assumed that the BS-zeRIS distance {\small$d_1$} is set equal to 20 m, and the zeRIS-UE distance {\small$d_2$} is set equal to 40 m. As it can be observed, the analytical and numerical results coincide, which validates the accuracy of the derived analysis. Furthermore, it can be seen that a higher number of reflecting elements in the zeRIS plays a critical role in reducing the required power for EH. Specifically, the joint energy-data rate outage probability approaches unity for large values of {\small$\rho$}, indicating a data rate outage phenomenon. This behavior can be attributed to the fact that as {\small$\rho \rightarrow 1$}, the received SNR tends to zero, leading to degraded data rate performance. Therefore, Fig. \ref{fig:PBPS} highlights the importance of optimizing the design of the zeRIS, as well as the selection of {\small$\rho$} for efficient zeRIS-assisted wireless communication systems.

Figs. \ref{fig:PBΤS} and \ref{fig:ES} depict the joint energy data rate outage probability of a \textit{BS-side} zeRIS-assisted network for the cases where the zeRIS absorbs energy through the TS and ES methods, respectively. It can be observed that the TS method exhibits a range of {\small$\tau$} values that result in low outage probabilities, which becomes narrower as {\small$N$} decreases, while similarly, the ES method shows the same trend for the set of {\small$\nu=\sfrac{N_1}{N}$} values. Specifically, when {\small$N=225$}, the set of {\small$\tau$} values that yield favorable joint energy-data rate probabilities is considerably narrow for the TS method, while for the ES method, no value of the ratio {\small$\nu$} leads to satisfactory performance in terms of outage probability. Notably, the optimal values of {\small$\tau$} and {\small$\nu$} for joint energy-data rate outage probability vary with increasing {\small$N$}. This emphasizes the need to apply each HaR method in consideration of the available number of reflecting elements, {\small$N$}. Finally, it is noticeable that, in a similar manner to the PS method, as either {\small$\tau$} or {\small$\nu$} approaches 1, the resulting joint energy-data rate outage probability also becomes equal to unity.

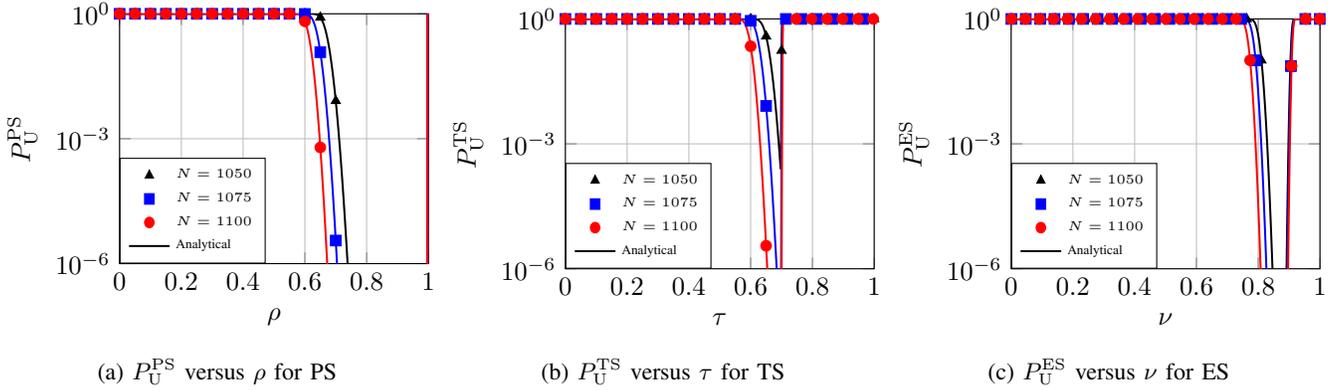
\begin{figure*}
\begin{subfigure}[t]{.32\textwidth}
	\centering
	\begin{tikzpicture}
	\begin{semilogyaxis}[
	width=0.98\linewidth,
	xlabel = {$\rho$},
	ylabel = {${P}_{\mathrm{U}}^{\mathrm{PS}} $},
	xmin = 0,xmax = 1,
	ymin = 0.000001,
	ymax = 1,
	xtick = {0,0.2,...,1},
	grid = major,
	legend image post style={xscale=0.9},
	legend cell align = {left},
      legend style={at={(0,0)},anchor=south west, font = \tiny}
	]
	\addplot[
	black,
      only marks,
	mark=triangle*,
	mark repeat = 50,
	mark size = 2,
	]
	table {data/PS_US_1050.dat};
 	\addlegendentry{$N=1050$}
	\addplot[
	blue,
        only marks,
	mark=square*,
	mark options={solid},
	mark repeat = 50,
	mark size = 2,
	]
	table {data/PS_US_1075.dat};	
	\addlegendentry{$N=1075$}
 	\addplot[
	red,
        only marks,
	mark=*,
	mark options={solid},
	mark repeat = 50,
	mark size = 2,
	]
	table {data/PS_US_1100.dat};	
	\addlegendentry{$N=1100$}
	\addplot[
	black,
       no marks,
	line width = 0.75pt,
	style = solid,
	]
	table {data/PS_US_1050.dat};
	\addlegendentry{Analytical}
	\addplot[
	blue,
        no marks,
	line width = 0.75pt,
	style = solid,
	]
	table {data/PS_US_1075.dat};	
 	\addplot[
	red,
        no marks,
	line width = 0.75pt,
	style = solid,
	]
	table {data/PS_US_1100.dat};	
	\end{semilogyaxis}
	\end{tikzpicture}
 \vspace{-0.3cm}
	\caption{${P}_{\mathrm{U}}^{\mathrm{PS}} $ versus $\rho$ for PS}
	\label{fig:UPS}
\end{subfigure}
\begin{subfigure}[t]{.32\textwidth}
	\centering
	\begin{tikzpicture}
	\begin{semilogyaxis}[
	width=0.98\linewidth,
	xlabel = {$\tau$},
	ylabel = {${P}_{\mathrm{U}}^{\mathrm{TS}} $},
	xmin = 0,xmax = 1,
	ymin = 0.000001,
	ymax = 1,
	xtick = {0,0.2,...,1},
	grid = major,
	legend image post style={xscale=0.9},
	legend cell align = {left},
      legend style={at={(0,0)},anchor=south west, font = \tiny}
	]
	\addplot[
	black,
      only marks,
	mark=triangle*,
	mark repeat = 50,
	mark size = 2,
	]
	table {data/TS_US_1050.dat};
 	\addlegendentry{$N=1050$}
	\addplot[
	blue,
        only marks,
	mark=square*,
	mark options={solid},
	mark repeat = 50,
	mark size = 2,
	]
	table {data/TS_US_1075.dat};	
	\addlegendentry{$N=1075$}
 	\addplot[
	red,
        only marks,
	mark=*,
	mark options={solid},
	mark repeat = 50,
	mark size = 2,
	]
	table {data/TS_US_1100.dat};	
	\addlegendentry{$N=1100$}
	\addplot[
	black,
       no marks,
	line width = 0.75pt,
	style = solid,
	]
	table {data/TS_US_1050.dat};
	\addlegendentry{Analytical}
	\addplot[
	blue,
        no marks,
	line width = 0.75pt,
	style = solid,
	]
	table {data/TS_US_1075.dat};	
 	\addplot[
	red,
        no marks,
	line width = 0.75pt,
	style = solid,
	]
	table {data/TS_US_1100.dat};	
	\end{semilogyaxis}
	\end{tikzpicture}
 \vspace{-0.3cm}
	\caption{${P}_{\mathrm{U}}^{\mathrm{TS}} $ versus $\tau$ for TS}
	\label{fig:UTS}
\end{subfigure}
\begin{subfigure}[t]{.32\textwidth}
	\centering
	\begin{tikzpicture}
	\begin{semilogyaxis}[
	width=0.98\linewidth,
	xlabel = {$\nu$},
	ylabel = {${P}_{\mathrm{U}}^{\mathrm{ES}} $},
	xmin = 0,xmax = 1,
	ymin = 0.000001,
	ymax = 1,
	xtick = {0,0.2,...,1},
	grid = major,
	legend image post style={xscale=0.9},
	legend cell align = {left},
      legend style={at={(0,0)},anchor=south west, font = \tiny}
	]
	\addplot[
	black,
      only marks,
	mark=triangle*,
	mark repeat = 50,
	mark size = 2,
	]
	table {data/ES_US_1050.dat};
 	\addlegendentry{$N=1050$}
	\addplot[
	blue,
        only marks,
	mark=square*,
	mark options={solid},
	mark repeat = 50,
	mark size = 2,
	]
	table {data/ES_US_1075.dat};	
	\addlegendentry{$N=1075$}
 	\addplot[
	red,
        only marks,
	mark=*,
	mark options={solid},
	mark repeat = 50,
	mark size = 2,
	]
	table {data/ES_US_1100.dat};	
	\addlegendentry{$N=1100$}
	\addplot[
	black,
       no marks,
	line width = 0.75pt,
	style = solid,
	]
	table {data/ES_US_1050.dat};
	\addlegendentry{Analytical}
	\addplot[
	blue,
        no marks,
	line width = 0.75pt,
	style = solid,
	]
	table {data/ES_US_1075.dat};	
 	\addplot[
	red,
        no marks,
	line width = 0.75pt,
	style = solid,
	]
	table {data/ES_US_1100.dat};	
	\end{semilogyaxis}
	\end{tikzpicture}
 \vspace{-0.3cm}
	\caption{${P}_{\mathrm{U}}^{\mathrm{ES}} $ versus $\nu$ for ES}
	\label{fig:UES}
\end{subfigure}
\caption{Joint energy-data rate outage probability for \textit{UE-side} zeRIS with {\small$R_{\mathrm{thr}}$}={\small$3.46 $} bit/s/Hz}
\end{figure*}

Fig. \ref{fig:UPS} presents the effect of the PS factor {\small$ \rho $} on the joint energy-data rate outage probability of a \textit{UE-side} zeRIS-assisted communication network. It is worth noting that in the \textit{UE-side} zeRIS case, {\small$ d_1 $} is set to 40 m, while {\small$ d_2 $} is set to 20 m. Similarly with the \textit{BS-side} zeRIS case, increasing the number of reflecting elements {\small$ N $} expands the set of {\small$ \rho $} values that result in optimal performance. However, to achieve low outage probability, {\small$ N $} and {\small$ \rho $} need to be larger compared to the \textit{BS-side} zeRIS case. This implies that the performance of a \textit{UE-side} zeRIS-assisted network is not symmetric to that of a \textit{BS-side} zeRIS-assisted network, even when the system parameters are set to be the same in both cases, which does not hold for a single RIS-assisted network. Furthermore, it is noteworthy that the range of {\small$ \rho $} values that enable low outage probability is narrower in the \textit{UE-side} zeRIS case compared to the \textit{BS-side} zeRIS case. For instance, when {\small$ N=1100 $} in the \textit{UE-side} zeRIS case, to achieve an outage probability lower than {\small$ 10^{-6} $}, {\small$ \rho $} must fall within the range {\small$ \left[0.68, 0.99\right] $}, while in the case of {\small$ N=275 $} in the \textit{BS-side} zeRIS case, {\small$ \rho $} must be within the range {\small$ \left[0.41, 0.93\right] $}. Hence, despite the fact that {\small$ N $} is four times larger in the \textit{UE-side} scenario compared to the \textit{BS-side} case, the range of {\small$ \rho $} values is narrower. This highlights the importance of carefully selecting the number of reflecting elements {\small$ N $} and the PS factor {\small$ \rho $} based on the specific zeRIS deployment scenario.

Figs. \ref{fig:UTS} and \ref{fig:UES} depict the joint outage probability of energy and data rate for the \textit{UE-side} zeRIS case, when the TS and ES methods are applied. The results indicate that a higher number of reflecting elements {\small$N$} is required in both the TS and ES methods for the \textit{UE-side} zeRIS case compared to the \textit{BS-side} zeRIS case, to achieve a low joint energy-data rate outage probability. Furthermore, the analysis reveals that increasing the value of {\small$N$} expands the range of values for the parameters {\small$\tau$} (for the TS method) and {\small$\nu$} (for the ES method) that result in a low outage probability. Notably, even by selecting the smallest examined value of {\small$N$}, i.e., {\small$N=1050$}, the TS method fails to achieve a joint energy-data rate outage probability lower than {\small$10^{-6}$}, unlike the ES method. Conversely, in the \textit{BS-side} case, by selecting the smallest examined value of {\small$N$}, i.e., {\small$N=225$}, there exists a small range of {\small$\tau$} values that can achieve low joint energy-data rate outage probability, while for the ES method, there is no value of {\small$\nu$} that can achieve the same. Therefore, it can be concluded that the number of zeRIS elements as well as the zeRIS location affect the optimal HaR method.

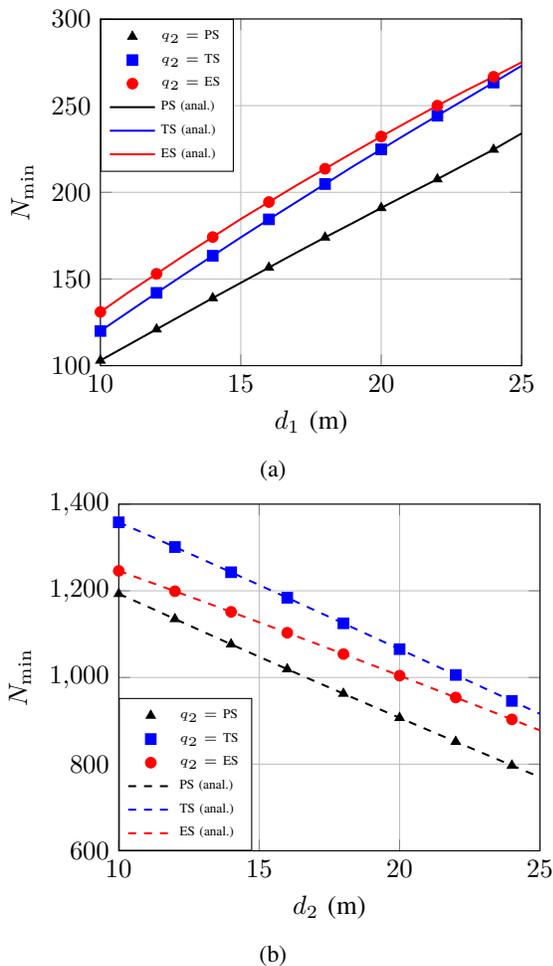
\begin{figure}[h!]
    \centering
    \begin{subfigure}{.4\textwidth}
        \centering
        \begin{tikzpicture}
	\begin{axis}[
	width=0.99\linewidth,
	xlabel = {$d_1$ (m)},
	ylabel = {$N_{\mathrm{min}}$},
	xmin = 10,xmax = 25,
	ymin = 100,
	ymax = 300,
	xtick = {10,15,...,25},
	grid = major,
	legend cell align = {left},
      legend style={at={(0,1)},anchor=north west, font = \tiny}
	]
	\addplot[
	black,
      only marks,
	mark=triangle*,
	mark repeat = 2,
	mark size = 2,
	]
	table {data/PS_BS_Nmin.dat};
 	\addlegendentry{$q_2=\text{PS}$}
	\addplot[
	blue,
        only marks,
	mark=square*,
	mark options={solid},
	mark repeat = 2,
	mark size = 2,
	]
	table {data/TS_BS_Nmin.dat};	
 	\addlegendentry{$q_2=\text{TS}$}
 	\addplot[
	red,
        only marks,
	mark=*,
	mark options={solid},
	mark repeat = 2,
	mark size = 2,
	]
	table {data/ES_BS_Nmin.dat};	
 	\addlegendentry{$q_2=\text{ES}$}
	\addplot[
	black,
       no marks,
	line width = 0.75pt,
	style = solid,
	]
	table {data/PS_BS_Nmin.dat};
        \addlegendentry{PS (anal.)}
	\addplot[
	blue,
        no marks,
	line width = 0.75pt,
	style = solid,
	]
	table {data/TS_BS_Nmin.dat};	
        \addlegendentry{TS (anal.)}
 	\addplot[
	red,
        no marks,
	line width = 0.75pt,
	style = solid,
	]
	table 
        {data/ES_BS_Nmin.dat};	
        \addlegendentry{ES (anal.)}
	\end{axis}
        \end{tikzpicture}
        \caption{}
        \label{fig:BSNmin}
    \end{subfigure}%

    \begin{subfigure}{.4\textwidth}
        \centering
        \begin{tikzpicture}
 	\begin{axis}[
	width=0.99\linewidth,
	xlabel = {$d_2$ (m)},
	ylabel = {$N_{\mathrm{min}}$},
	xmin = 10,xmax = 25,
	ymin = 600,
	ymax = 1400,
	xtick = {10,15,...,25},
	grid = major,
	legend cell align = {left},
      legend style={at={(0,0)},anchor=south west, font = \tiny}
	]
	\addplot[
	black,
      only marks,
	mark=triangle*,
	mark repeat = 2,
	mark size = 2,
	]
	table {data/PS_US_Nmin.dat};
 	\addlegendentry{$q_2=\text{PS}$}
	\addplot[
	blue,
        only marks,
	mark=square*,
	mark options={solid},
	mark repeat = 2,
	mark size = 2,
	]
	table {data/TS_US_Nmin.dat};	
 	\addlegendentry{$q_2=\text{TS}$}
 	\addplot[
	red,
        only marks,
	mark=*,
	mark options={solid},
	mark repeat = 2,
	mark size = 2,
	]
	table {data/ES_US_Nmin.dat};	
 	\addlegendentry{$q_2=\text{ES}$}
	\addplot[
	black,
       no marks,
	line width = 0.75pt,
	style = dashed,
	]
	table {data/PS_US_Nmin.dat};
        \addlegendentry{PS (anal.)}
	\addplot[
	blue,
        no marks,
	line width = 0.75pt,
	style = dashed,
	]
	table {data/TS_US_Nmin.dat};	
        \addlegendentry{TS (anal.)}
 	\addplot[
	red,
        no marks,
	line width = 0.75pt,
	style = dashed,
	]
	table 
        {data/ES_US_Nmin.dat};	
        \addlegendentry{ES (anal.)}
	\end{axis}
        \end{tikzpicture}
        \caption{}
        \label{fig:USNmin}
    \end{subfigure}
    \caption{Analysis of minimum reflecting elements for: (a) \textit{BS-side} zeRIS with $d_2=60-d_1$ and $R_{\mathrm{thr}}=1$ bit/s/Hz, (b) \textit{UE-side} zeRIS with $d_1=60-d_2$ and $R_{\mathrm{thr}}=1$ bit/s/Hz.}
    \label{fig:Nminanalysis}
\end{figure}

Fig. \ref{fig:Nminanalysis} illustrates the minimum required number of reflecting elements, $N_{\mathrm{min}}$, to achieve a joint energy-data rate outage probability lower than or equal to $10^{-6}$ in both \textit{BS-side} and \textit{UE-side} zeRIS deployment scenarios, for the case where the rate threshold $R_{\mathrm{thr}}=3.46$ bit/s/Hz. Specifically, in both \textit{BS-side} and \textit{UE-side} setups, the PS method consistently stands out by requiring the fewest reflecting elements than both the TS and ES methods. In addition, an interesting observation in both cases is the inverse relationship between the BS-zeRIS distance and $N_{\mathrm{min}}$, emphasizing the paramount importance of the EH link over the communication link in zeRIS operation. Additionally, the performance asymmetry between \textit{UE-side} and \textit{BS-side} zeRIS-assisted networks is evident, with the former requiring a significantly higher number of reflecting elements. Finally, Fig. 4a and Fig. 4b highlight a widening gap in $N_{\mathrm{min}}$ between PS and TS and between PS and ES, respectively,  while the difference between TS and ES in both figures narrows, indicating potential scenarios where one method might require fewer reflecting elements.

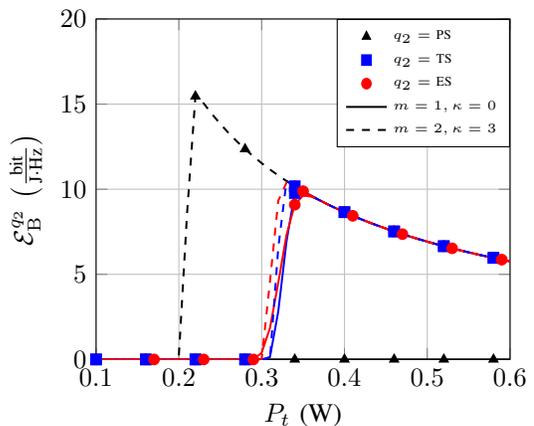
\begin{figure}
	\centering
	\begin{tikzpicture}
	\begin{axis}[
	width=0.8\linewidth,
	xlabel = {$P_t$ (W)},
	ylabel = {${\mathcal{E}_{\mathrm{B}}^{q_2}}$ $\left(\frac{\text{bit}}{\text{J}\cdot\text{Hz}}\right)$},
	xmin = 0.1,xmax = 0.6,
	ymin = 0,
	ymax = 20,
	xtick = {0,0.1,...,0.6},
	grid = major,
	legend image post style={xscale=0.9},
	legend cell align = {left},
      legend style={at={(1,1)},anchor=north east, font = \tiny}
	]
	\addplot[
	black,
      only marks,
	mark=triangle*,
	mark repeat = 6,
	mark size = 2,
	]
	table {data/PS_BS_comp_10_Rayleigh.dat};
 	\addlegendentry{$q_2=\text{PS}$}
	\addplot[
	blue,
        only marks,
	mark=square*,
	mark options={solid},
	mark repeat = 6,
	mark size = 2,
	]
	table {data/TS_BS_comp_10_Rayleigh.dat};	
 	\addlegendentry{$q_2=\text{TS}$}
 	\addplot[
	red,
        only marks,
	mark=*,
	mark options={solid},
	mark repeat = 6,
	mark size = 2,
	]
	table {data/ES_BS_comp_10_Rayleigh.dat};	
 	\addlegendentry{$q_2=\text{ES}$}
	\addplot[
	black,
       no marks,
	line width = 0.75pt,
	style = solid,
	]
	table {data/PS_BS_comp_10_Rayleigh.dat};
    \addlegendentry{$m=1,\kappa=0$}
    	\addplot[
	black,
       no marks,
	line width = 0.75pt,
	style = dashed,
	]
	table {data/PS_BS_comp_10.dat};
	\addlegendentry{$m=2, \kappa=3$}
	\addplot[
	blue,
        no marks,
	line width = 0.75pt,
	style = solid,
	]
	table {data/TS_BS_comp_10_Rayleigh.dat};	
 	\addplot[
	red,
        no marks,
	line width = 0.75pt,
	style = solid,
	]
	table {data/ES_BS_comp_10_Rayleigh.dat};	
	\addplot[
	black,
      only marks,
	mark=triangle*,
	mark repeat = 6,
	mark size = 2,
	]
	table {data/PS_BS_comp_10.dat};
	\addplot[
	blue,
        only marks,
	mark=square*,
	mark options={solid},
	mark repeat = 6,
	mark size = 2,
	]
	table {data/TS_BS_comp_10.dat};	
 	\addplot[
	red,
        only marks,
	mark=*,
	mark options={solid},
	mark repeat = 6,
	mark size = 2,
        mark phase = 8.5,
	]
	table {data/ES_BS_comp_10.dat};	
	\addplot[
	blue,
        no marks,
	line width = 0.75pt,
	style = dashed,
	]
	table {data/TS_BS_comp_10.dat};	
 	\addplot[
	red,
        no marks,
	line width = 0.75pt,
	style = dashed,
	]
	table {data/ES_BS_comp_10.dat};
	\end{axis}
	\end{tikzpicture}
 \vspace{-0.3cm}
	\caption{Energy efficiency versus transmit power for a \textit{BS-side} zeRIS with $N=275$ and $R_{\mathrm{thr}}=3.46$ bit/s/Hz.}
	\label{fig:EEBS}
\end{figure}

In Fig. \ref{fig:EEBS}, we examine the normalized energy efficiency {\small$\mathcal{E}_{\mathrm{B}}^{q_2}$} of a \textit{BS-side} zeRIS-assisted network, with respect to the transmit power {\small$P_t$}. For this analysis, we consider a \textit{BS-side} zeRIS configuration equipped with 275 reflecting elements and a rate threshold set at {\small$R_{\mathrm{thr}}=3.46$} bit/s/Hz. We delve into two distinct propagation conditions: i) a scenario with LoS component described by the parameters {\small$m=2$} and {\small$\kappa=3$}, and ii) a non-LoS scenario, where {\small$m=1$} and {\small$\kappa=0$}. Our primary objective is to identify the most appropriate HaR method that effectively balances power efficiency and reliable communication. To do so, we determine the optimal PS factor {\small$\rho$}, time splitting factor {\small$\tau$}, and the proportion of elements allocated for energy harvesting {\small$\frac{N_1}{N}$} for each {\small$P_t$} value, aiming to maximize energy efficiency. As it can be observed, while the PS method stands out for energy efficiency in the LoS scenario, it does not perform as well in NLoS situations. In contrast, both the TS and ES methods showcase consistent performance across different link conditions. Particularly, the efficiency of the PS method within the \textit{BS-side} zeRIS context is significantly influenced by propagation conditions, highlighting the need to consider the propagation environment when selecting a HaR method.

\begin{figure}
	\centering
	\begin{tikzpicture}
	\begin{axis}[
	width=0.8\linewidth,
	xlabel = {$P_t$ (W)},
	ylabel = {${\mathcal{E}_{\mathrm{U}}^{q_2}} \left(\frac{\text{bit}}{\text{J}\cdot\text{Hz}}\right)$},
	xmin = 0.1,xmax = 0.6,
	ymin = 0,
	ymax = 18,
	xtick = {0,0.1,...,0.6},
	grid = major,
	legend image post style={xscale=0.9},
	legend cell align = {left},
      legend style={at={(1,1)},anchor=north east, font = \tiny}
	]
	\addplot[
	black,
      only marks,
	mark=triangle*,
	mark repeat = 2,
	mark size = 2,
	]
	table {data/PS_US_comp_10.dat};
 	\addlegendentry{$q_2=\text{PS}$}
	\addplot[
	blue,
        only marks,
	mark=square*,
	mark options={solid},
	mark repeat = 2,
	mark size = 2,
	]
	table {data/TS_US_comp_10.dat};	
 	\addlegendentry{$q_2=\text{TS}$}
 	\addplot[
	red,
        only marks,
	mark=*,
	mark options={solid},
	mark repeat = 2,
	mark size = 2,
	]
	table {data/ES_US_comp_10.dat};	
 	\addlegendentry{$q_2=\text{ES}$}
	\addplot[
	black,
       no marks,
	line width = 0.75pt,
	style = solid,
	]
	table {data/PS_US_comp_10.dat};
    \addlegendentry{$R_{\mathrm{thr}}=3.46$ (anl.)}
    \addplot[
	black,
       no marks,
	line width = 0.75pt,
	style = dashed,
	]
	table {data/PS_US_comp_1.dat};
	\addlegendentry{$R_{\mathrm{thr}}=1$ (anl.)}
	\addplot[
	blue,
        no marks,
	line width = 0.75pt,
	style = solid,
	]
	table {data/TS_US_comp_10.dat};	
 	\addplot[
	red,
        no marks,
	line width = 0.75pt,
	style = solid,
	]
	table {data/ES_US_comp_10.dat};	
	\addplot[
	black,
      only marks,
	mark=triangle*,
	mark repeat = 4,
	mark size = 2,
        mark phase = 14,
	]
	table {data/PS_US_comp_1.dat};
	\addplot[
	blue,
        only marks,
	mark=square*,
	mark options={solid},
	mark repeat = 4,
	mark size = 2,
        mark phase = 8,
	]
	table {data/TS_US_comp_1.dat};	
 	\addplot[
	red,
        only marks,
	mark=*,
	mark options={solid},
	mark repeat = 4,
	mark size = 2,
        mark phase = 8.5,
	]
	table {data/ES_US_comp_1.dat};	

	\addplot[
	blue,
        no marks,
	line width = 0.75pt,
	style = dashed,
	]
	table {data/TS_US_comp_1.dat};	
 	\addplot[
	red,
        no marks,
	line width = 0.75pt,
	style = dashed,
	]
	table {data/ES_US_comp_1.dat};
	\end{axis}
	\end{tikzpicture}
 \vspace{-0.3cm}
	\caption{Energy efficiency versus transmit power for a \textit{UE-side} zeRIS with $N=1100$}
	\label{fig:EEUS}
\end{figure}
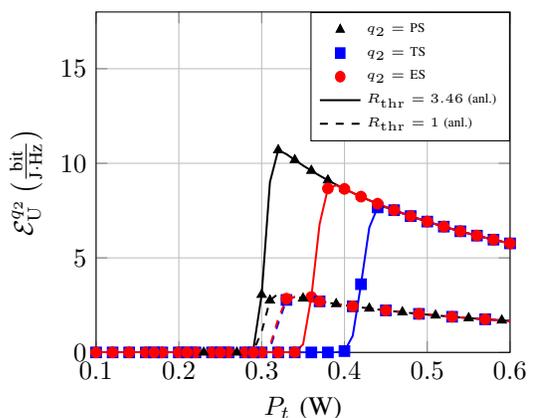

Finally, accordingly to the \textit{BS-side} zeRIS case, Fig. \ref{fig:EEUS} depicts the normalized energy efficiency of the \textit{UE-side} zeRIS-assisted system as a function of the transmit power {\small$P_t$} for two $R_{\mathrm{thr}}=3.46$ cases where i) {\small$R_{\mathrm{thr}}=3.46$} bit/s/Hz, and ii) {\small$R_{\mathrm{thr}}=1$} bit/s/Hz. Initially, it can be observed that a \textit{UE-side} zeRIS requires more reflecting elements than a \textit{BS-side} zeRIS, which highlights the fact that the performance of a \textit{UE-side} zeRIS-assisted system is not symmetric with its \textit{BS-side} equivalent, in contrast to a conventional RIS-assisted system. Specifically, for the case where {\small$R_{\mathrm{thr}}=3.46$} bit/s/Hz, the ES method outperforms TS which contradicts the \textit{BS-side} zeRIS case, while the PS method illustrates the best performance among the examined HaR methods. Finally, for the case where {\small$R_{\mathrm{thr}}=1$} bit/s/Hz, the ES method has the same performance as the TS method which is slightly worse than PS. Finally, it becomes evident that the choice of $R_{\mathrm{thr}}$ in various scenarios can significantly influence the selection of the most suitable HaR method.

\section{Conclusions}
In this research, we delved into zeRISs' capabilities to elevate energy-efficient communications for upcoming wireless networks and provided closed-form expressions for the joint energy-data rate outage probability and energy efficiency in both \textit{BS-side} and \textit{UE-side} zeRIS deployments. These expressions are instrumental in analyzing the performance of zeRIS-assisted networks. Emphasizing the significance of informed HaR method choices, our results showcase that zeRIS-assisted systems can be adeptly tailored to meet varied application demands, prioritizing energy conservation. Using PS, TS, and ES as HaR options, we pinpointed the vital roles of reflecting element count and zeRIS-HaR method selection in performance enhancement. The optimal HaR strategy leans on zeRIS placement, UE's target rate, and the propagation environment, and while PS often leads in energy efficiency, the balance between PS, TS, and ES is pivotal for system-specific conditions. Differing from conventional RIS systems, the performance of \textit{UE-side} zeRIS doesn't mirror its \textit{BS-side} counterpart, even with matching parameters. As a future direction, exploring the integration of active RISs, given their unique energy characteristics into the zeRIS paradigm presents a compelling field for research. 

\vspace{-1em}
\appendices
\section{Proof of Proposition \ref{prop:1}}
Considering equations \eqref{RPS}, \eqref{EPS}, \eqref{QPS}, and that {\small$\lvert h_{1i}\rvert=1$}, the joint energy-data rate outage probability for a \textit{BS-side} zeRIS-assisted network that applies the PS method can be expressed as 
\begin{equation}\label{PJBS1}
\small
\begin{split}
    &{P^{\mathrm{PS}}_{\mathrm{B}}} = \Pr \Bigg( \rho \zeta  P_t G_t \ell_1 \left| \sum_{i=1}^N  e^{j \mathrm{arg}(h_{2i})}\right|^2 \leq N P_e + P_{\mathrm{circ}} \\
    &\cup \ \mathrm{log_2}\Bigg(1 + \gamma_t G \ell \left(1- \rho \right)  \left| \sum_{i=1}^{N} \lvert h_{2i} \rvert \right|^2\Bigg) \leq R_{\mathrm{thr}} \Bigg).
\end{split}
\end{equation}
As it can be observed, the RVs that are upper-bounded in \eqref{PJBS1} are different and independent from each other, due to the fact that the zeRIS-UE channel's phase {\small$\mathrm{arg}(h_{2i})$} is a circular RV that is independent and not identically distributed with the channel gain {\small$\lvert h_{2i} \rvert$}, which is assumed to be a Nakagami-{\small$m$} RV. In more detail, by taking into account the model proposed in \cite{Abdi}, the phase of a channel that describes a wireless link with both LoS and non-LoS components (e.g., Nakagami-{\small$m$} channel), can be described as an RV with a probability density function (PDF) equal to
\begin{equation}
\small
    f_p(\theta) = \frac{e^{\kappa \mathrm{cos(\theta)}}}{2 \pi \left(K+1\right) I_0(\kappa)} + \frac{K}{K+1} \delta(\theta),
\end{equation}
where {\small$K$} is the Rice factor, which characterizes the ratio of the power in the LoS component to that in the non-LoS components and can be approximated as in \cite{kappa}
\begin{equation}
\small
    K\approx \frac{\sqrt{m^2 - m}}{m - \sqrt{m^2-m}}.
\end{equation}
Additionally, {\small$\delta(\cdot)$} is the Dirac function, and {\small$kappa$} is the von Mises concentration parameter, which characterizes the width of the interval of phase values associated with the arriving signals, and describes the effect of the propagation environment on the arriving phases at the UE. Therefore, \eqref{PJBS1} can be rewritten as
\begin{equation} \label{PJBS2}
\small
\begin{split}
    {P^{\mathrm{PS}}_{\mathrm{B}}} &= \Pr \left( \left| \sum_{i=1}^N  e^{j \mathrm{arg}(h_{2i})}\right| \leq w_1\right) + \Pr \left( \left| \sum_{i=1}^{N} \lvert h_{2i} \rvert\right| \leq w_2 \right) \\
    & - \Pr \left( \left| \sum_{i=1}^N  e^{j \mathrm{arg}(h_{2i})}\right| \leq w_1\right) \Pr \left( \left| \sum_{i=1}^{N} \lvert h_{2i} \rvert\right| \leq w_2 \right).
\end{split}    
\end{equation}
Considering that {\small$N>50$} in order to enable the reliable performance of the different zeRIS electromagnetic functionalities, according to the results provided in \cite{justin}, {\small$\left| \sum_{i=1}^N  e^{j \mathrm{arg}(h_{2i})}\right|$} can be efficiently approximated by a folded normal RV with mean value {\small$\mu= N \phi_1$}, and variance {\small$\sigma^2=\frac{N}{2} \left(1 + \phi_2 -2 \phi_1^2\right)$}, where {\small$\phi_i$} is the {\small$i$}-th trigonometric moment of {\small$\mathrm{arg}(h_{2i})$}, and can be calculated as
\begin{equation}\label{trig}
\small
    \phi_n = \int_{0}^{2\pi} e^{jn\theta} f_p(\theta) d\theta.
\end{equation}
Therefore, after some algebraic manipulations, the first and the second trigonometric moments of {\small$\mathrm{arg}(h_{2i})$} are derived as
\begin{equation}
\small
    \phi_1 = \frac{I_1(\kappa)}{I_0(\kappa)\left(K+1\right)} + \frac{K}{K+1}
\end{equation}
 and
\begin{equation}
\small
    \phi_2 = \frac{I_2(\kappa)}{I_0(\kappa)\left(K+1\right)} + \frac{K}{K+1}.
\end{equation}
Moreover, by employing the moment-matching technique, the RV {\small$Z_1=\left| \sum_{i=1}^{N} \lvert h_{2i} \rvert\right|$} can be accurately approximated by a gamma-distributed RV with scale parameter {\small$k_{\mathrm{PS}}=\frac{\mathbb\mathbb{E}^2[Z_1]}{\mathrm{Var}[Z_1]}$} and shape parameter {\small$\theta_{\mathrm{PS}}=\frac{\mathrm{Var}[Z_1]}{\mathbb{E}[Z_1]}$}, where {\small$\mathrm{Var}[\cdot]$} denotes variance. Thus, we need to calculate the mean value and the variance of {\small$Z_1$} which can be expressed, respectively, as
\begin{equation}
\small
\begin{split}
    \mathbb{E}[Z]= \mathbb{E}\left[\left| \sum_{i=1}^N \left| h_{2i}\right| \right|\right]=N \mathbb{E}\left[\left|h_{2i}\right|\right]= N \sqrt{\frac{\Omega}{m}}\frac{\Gamma(m+\frac{1}{2})}{\Gamma(m)},
\end{split}
\end{equation}
and
\begin{equation}
\small
\begin{split}
    \mathrm{Var}[Z]&= \mathrm{Var}\left[\left| \sum_{i=1}^N \left| h_{2i}\right| \right|\right]=N \mathrm{Var}\left[\left|h_{2i}\right|\right]=\\
    &= N \Omega \left(1 - \frac{1}{m}\left(\frac{\Gamma(m+\frac{1}{2})}{\Gamma(m)}\right)^2 \right).
\end{split}
\end{equation}
Thus, by utilizing the CDFs of the folded normal distribution and the gamma distribution, which are equal to
\begin{equation}
\small
    F_f(x) = \frac{1}{2} \left[\mathrm{erf}\left(\frac{x +\mu}{\sqrt{2 \sigma^2}} \right) +  \mathrm{erf}\left(\frac{x -\mu}{\sqrt{2 \sigma^2}} \right)\right],
\end{equation}
and
\begin{equation}\label{CDF}
\small
    F_g(x) = \frac{\gamma(k_{\mathrm{PS}},\frac{x}{\theta_{\mathrm{PS}}} )}{\Gamma(k_{\mathrm{PS}})},
\end{equation}
respectively, then {\small${P^{\mathrm{PS}}_{j,B}}$} can be derived, which concludes the proof.

\bibliographystyle{IEEEtran}
\bibliography{Bibliography}

\end{document}